\documentstyle[aas2pp4]{article}

\def\ltsima{$\; \buildrel < \over \sim \;$}
\def\simlt{\lower.5ex\hbox{\ltsima}}            
\def\gtsima{$\; \buildrel > \over \sim \;$}
\def\simgt{\lower.5ex\hbox{\gtsima}}            

\newcommand{\asca}{{\it ASCA}}
\newcommand{\rosat}{{\it ROSAT}}
\newcommand{\einstein}{{\it Einstein}}
\newcommand{\ginga}{{\it Ginga}}
\newcommand{\heao}{{\it HEAO1}}

\newcommand{\logn}{Log $N$ - Log $S$ relation}
\newcommand{\Logn}{Log $N$ - Log $S$ Relation}
\newcommand{\etal}{{\it et al.}}

\newcommand{\ergs}{erg s$^{-1}$ cm$^{-2}$}
\newcommand{\ergss}{erg s$^{-1}$ cm$^{-2}$ sr$^{-1}$}
\newcommand{\de}{deg$^2$}

\newcommand{\nh}{$N_{\rm H}$}

\slugcomment{Accepted for publication in ApJ}
\lefthead{Ueda et al.}
\righthead{ASCA Large Sky Survey}

\begin{document}

\title{Log $N$ - Log $S$ Relations 
and Spectral Properties of Sources
from the ASCA Large Sky Survey
---
their Implications for the Origin of the Cosmic X-ray Background (CXB)
}

\author{Yoshihiro Ueda, Tadayuki Takahashi, and Hajime Inoue}
\affil{Institute of Space and Astronautical Science,
 Kanagawa 229-8510, Japan\\
 ueda@astro.isas.ac.jp, takahasi@astro.isas.ac.jp, inoue@astro.isas.ac.jp}

\author{Takeshi Tsuru and Masaaki Sakano}
\affil{Department of Physics, Kyoto University,
 Kyoto 606-8502, Japan\\
 tsuru@cr.scphys.kyoto-u.ac.jp, sakano@cr.scphys.kyoto-u.ac.jp}

\author{Yoshitaka Ishisaki}
\affil{Department of Physics, Tokyo Metropolitan University,
 Hachioji, Tokyo 192-0397, Japan\\
 ishisaki@phys.metro-u.ac.jp}

\author{Yasushi Ogasaka}
\affil{Department of Astrophysics, Nagoya University,
 Nagoya 464-8602, Japan\\
 ogasaka@u.phys.nagoya-u.ac.jp}

\author{Kazuo Makishima}
\affil{Department of Physics, University of Tokyo,
 Bunkyo-ku, Tokyo 113-0033, Japan\\
 maxima@phys.s.u-tokyo.ac.jp}

\author{Toru Yamada}
\affil{Astronomical Institute, Tohoku University,
 Sendai 980-8578, Japan\\
 yamada@astr.tohoku.ac.jp}

\and

\author{Masayuki Akiyama and Kouji Ohta}
\affil{Department of Astronomy, Kyoto University, 
 Kyoto 606-8502, Japan\\
 akiyama@kusastro.kyoto-u.ac.jp, ohta@kusastro.kyoto-u.ac.jp}

\newpage
\begin{abstract}

We carried out the first wide-area unbiased survey with the \asca\
satellite in the 0.7--10 keV band around a north Galactic-pole region
covering a continuous area of 7 deg$^2$ (Large Sky Survey; LSS). To
make the best use of \asca\ capability, we developed a new
source-detection method where the complicated detector responses are
fully taken into account. Applying this method to the entire LSS data
independently in the total (0.7--7 keV), hard (2--10 keV), and soft
(0.7--2 keV) band, we detected 107 sources altogether with sensitivity
limits of $6\times 10^{-14}$ (0.7--7 keV), $1\times 10^{-13}$ (2--10
keV), and $2\times 10^{-14}$ \ergs\ (0.7--2 keV), respectively. A
complete list of the detected sources is presented.  Based on detailed
studies by Monte Carlo simulations, we evaluated effects of the source
confusion and accurately derived \logn\ in each survey band. The
\logn\ in the hard band is located on the extrapolation from the
\ginga\ and \heao\ results with the Euclidean slope of $-3/2$, while
that in the soft band is consistent with the results by
\rosat. 
At these flux limits, 30($\pm$3)\% of the CXB in the 0.7--7 keV band and
23($\pm$3)\% in the 2--10 keV band have been resolved into discrete
sources. The average spectrum of faint sources detected in the total
band shows a photon index of 1.63$\pm$0.07 in the 0.7--10 keV range,
consistent with the comparison of source counts between the hard and
the soft energy band. Those detected in the hard band show a photon
index of 1.49$\pm$0.10 in the 2--10 keV range. These spectral properties
suggest that contribution of sources with hard energy spectra become
significant at a flux of $10^{-13}$ \ergs\ (2--10 keV).  The most
plausible candidates are type-II AGNs, as indicated by
on-going optical identifications.

\end{abstract}


\twocolumn

\section{Introduction}

Since the discovery of the Cosmic X-ray Background (CXB) more than 35
years ago (Giacconi \etal\ 1962), its origin has been a key issue of
the X-ray astronomy (for reviews see Fabian and Barcons 1992; Hasinger
1996).  Although the emission extends over a wide energy range in more
than 5 decades from X-rays to $\gamma$-rays, the main energy flux of
the CXB is emitted in the hard X-ray band, peaked at $\sim$ 20 keV.
The spectrum resembles that of a hot plasma with a temperature of 40
keV (Marshall \etal\ 1980; Rothschild \etal\ 1983), and this
introduced the possibility that hot plasmas uniformly fill the
universe. However, the low level of distortion of the Cosmic Microwave
Background (CMB) spectrum by the Compton up-scattering measured with
{\it COBE} ruled out this possibility, implying that the origin of the
CXB is the superposition of faint discrete sources (Mather \etal\
1994).

The understanding of the CXB origin has been quite different for the
soft ($<$2--3 keV), and the hard ($>$2--3 keV) energy bands,
respectively. In the soft X-ray band, the \rosat\ satellite resolved
70--80\% of the CXB in the 0.5--2 keV band into individual sources
(Hasinger \etal\ 1998). Most of them are optically identified as (type-I) AGNs
such as quasars (Schmidt \etal\ 1998). The average spectrum of these
AGNs coincides with that of the CXB in the 1--2 keV range having a
photon index of $\Gamma$ of $\simeq$ 2. The slope is steeper than that
of the CXB extrapolated from above 2 keV, $\Gamma \simeq$ 1.4 (e.g.,
Gendreau \etal\ 1995; Ishisaki \etal\ 1999).

By contrast, in the hard X-ray band above 2 keV, the origin of the CXB
has been less well understood, due to lack of imaging instruments.
{\it Uhuru} (Forman \etal\ 1978), {\it Ariel V} (McHardy \etal\ 1981),
\heao\ (Piccinotti \etal\ 1982), and \ginga\ (Kondo
\etal\ 1992) performed X-ray surveys above the 2 keV band. However,
since they carried collimated-type detectors where the source
confusion becomes significant, the achieved sensitivity limits are at
most $\sim 10^{-11}$ \ergs\ (2--10 keV), and the sources observed by
them only account for 3\% of the total emission of the CXB in the
2--10 keV band. Moreover, for the hard X-ray background,
there is a problem called the spectral paradox: bright AGNs observed
with \heao, {\it EXOSAT} and \ginga\ have spectra with an average
photon index $\Gamma$ of 1.7$-$1.9 (e.g., Turner and Pounds 1989;
Williams \etal\ 1992), which is significantly softer than that of the
CXB itself, and hence, cannot be responsible for the total CXB
emission. The spectra of these bright AGNs are consistent with the
spectrum of the CXB fluctuation observed with \ginga, which shows
$\Gamma$ = 1.8$\pm$0.1 in the 2--10 keV range (Hayashida, Inoue, \&
Kii 1990; Butcher \etal\ 1997).  Besides the spectral paradox, a
discrepancy of source counts between the soft and hard energy bands
has been noticed (e.g., Fabian \& Barcons 1992). The normalization of
the \logn\ in the 2--10 keV band obtained by the fluctuation analysis with
\heao\ (Shafer and Fabian 1983) and \ginga\ (Butcher \etal\ 1997) is
about 2 times larger than that obtained with the \einstein\ Medium
Sensitivity Survey (MSS) in the 0.3--3.5 keV band (Gioia \etal\ 1984;
1990), if we convert the flux assuming a photon index of $1.7$.

In order to study the nature of the X-ray source population over the
0.7--10 keV band, and to solve these unresolved problems on the CXB
origin, we performed an unbiased imaging sky survey with the \asca\
satellite (Large Sky Survey: LSS; Ueda 1996; Inoue \etal\ 1996).
\asca\ (Tanaka, Inoue, 
\& Holt 1994), carrying four identical X-Ray Telescopes (XRT;
Serlemitsos \etal\ 1995) coupled to two Solid-state Imaging
Spectrometers (SIS; Burke \etal\ 1991; Yamashita \etal\ 1997), and two
Gas Imaging Spectrometers (GIS; Ohashi \etal\ 1996; Makishima \etal\
1996) as focal plane detectors, is the first imaging satellite capable
of study of the X-ray band above 2 keV with a sensitivity up to
several $ 10^{-14}$
\ergs\ (2--10 keV). It also covers the wide energy band from 0.5 to 10
keV, which allows us to directly compare results of the energy bands
below and above 2 keV with single detectors, hence accompanied with
much less uncertainties than previous studies. The first results from
the LSS focusing on the 2--10 keV band data are reported by Ueda \etal\
(1998).

In this paper, we present the details of the analysis and the results,
 derived utilizing the whole data over the 0.7--10 keV band.  In section~2, we
summarize observations and data reduction.  In section~3, we describe
analysis methods and results. A complete X-ray source list in the LSS field
is presented. In section~4, we evaluate possible systematic errors in
our results, based on studies by Monte Carlo simulations. In section~5,
\logn s are derived in three survey bands (0.7--7 keV, 2--10 keV and
0.7--2 keV).  In section~6, we discuss implication of the results.

\section{Observations and Data Reduction}
\subsection{Observations}

The survey field of the LSS is a continuous region near the north
Galactic pole, centered at $RA$(2000) = 13$^{\rm h}$14$^{\rm m}$,
$DEC$(2000)=31$^\circ 30'$. We had no bias in selecting this field
except that we avoided bright X-ray sources with a flux greater than
$\sim 10^{-11}$ \ergs\ (2--10 keV) in the major X-ray source catalogs
in order to minimize complexity in the analysis that would be induced
by bright sources. The Galactic neutral hydrogen column density toward
the LSS field is $N_{\rm H} = 1.1\times 10^{20}$ cm$^{-2}$, which is
sufficiently low to study extra-Galactic sources above 0.7 keV.

The observation log of the LSS is summarized in Table~1. Seventy-six
pointings have been made in total (16 pointings in Dec.\ 1993 and
Jan.\ 1994, 20 pointings in June 1994, 20 pointings in Dec.\ 1994 and
Jan.\ 1995, and 20 pointings in June/July 1995) with a typical
exposure of 10 ksec per pointing. Shifting the pointing positions
by the half size of the SIS FOV in sequence, we observed each sky
point twice with the SIS so that the sum of field of views (FOVs) of
the SIS ($22'\times22'$) covered the survey area continuously with
good overlaps. This operation makes the sensitivity limits quite
uniform over the entire field. The total sky area observed with the GIS
and SIS amounts to 7.0 \de, and 5.4 \de, respectively. The mean
exposure time per point is 56 ksec (sum of GIS2 and GIS3), and 23 ksec
(sum of SIS0 and SIS1) after data screening.

\placetable{tbl-1}

\subsection{Data Reduction}

Since the LSS observations consist of many pointings, we created
images in the sky coordinate system (Gotthelf 1996) to be superposed
on a common coordinate system for all the data. Photons were accumulated in
the different energy bands of 0.7--7 keV, 2--7 keV, and 0.7--2 keV for
the SIS, and 0.7--7 keV, 2--10 keV, 0.7--7 keV, 2--4 keV, and 4--10
keV for the GIS so that we can get the spectral information from
energy-dependent image analysis. We did not use energies below 0.7
keV, nor energies above 10 keV (7 keV for the SIS), because the
contamination by the Non X-ray Background (NXB) is high relative to
X-ray events (e.g., Kubo \etal\ 1994). We utilized as much data as
possible, so that the Signal to Noise (S/N) ratio should be optimized
for faint source detection (Ueda 1996). We discarded data of the outer
part of the GIS images with a radius larger than 18$'$ from the
optical axis, which are not suitable for study of faint sources
because (1) the NXB rate is high (Kubo \etal\ 1994), (2) those data
are greatly affected by the stray light from the CXB outside the FOV
(Ishisaki 1996), and (3) the Point Spread Function (PSF) of the XRT
have not been directly measured through the observations of Cyg X-1
(Takahashi \etal\ 1995).

The nominal error of the attitude solution of
\asca\ is about 1 arcmin (Gotthelf 1995), and arises from the mis-alignment
between the focal plane detectors and attitudes sensors (Ueda \etal\
1999). The error depends on the temperature of the satellite
base-plate. In our analysis we corrected for this error when
converting from the detector coordinate to the sky coordinate, using
the onboard house-keeping data according to the method described in
Ueda \etal\ (1999). This reduces the systematic positional error, due
to the absolute attitude error, to 0.5 arcmin in radius (90\%
confidence level). Consequently, we could greatly improve the quality
of images superposed from the multi pointings with different orbit
conditions of the satellite. The final position uncertainties, which
contain statistical errors, are evaluated in the next section.

\section{Analysis and Results}

Extremely careful analysis is required to study faint sources 
with \asca.
First, the PSF of the XRT has a large scattering tail, which affects
flux determination of a faint source near a brighter one. The PSF
itself strongly depends on the position of the detector as well as
photon energy (Serlemitsos \etal\ 1995). The background, consisting of
the CXB and the NXB, also depends on the position and energy. It
should be noted that the intensity of the NXB can fluctuate by a
factor of 10--20\% from pointing to pointing (Ishisaki \etal\ 1997). 
Furthermore, there is a grid support on the GIS detector (Ohashi
\etal\ 1996), which should be taken into account in calculating fluxes
of sources falling near its structure. Under these conditions, we must
pay attention so that real sources are found completely but
statistical fluctuations are not picked up. Also, we should be very
careful in determining fluxes of faint sources, which are highly
sensitive to background subtraction under the low S/N ratio.

To overcome these difficulties in analysis of \asca\ data, we
developed a new source-detection method (Ishisaki \etal\ 1995; Ueda
1996; Takahashi \etal\ 1998). In this procedure we take fully account
of the complicated detector responses, such as the position-dependent
PSF, the background, the vignetting of the XRT, and the grid shadows
of the GIS. The procedure consists of two steps: (Step~I) detection of
source candidates and (Step~II) calculation of their statistical
significances and fluxes. A source candidate whose significance
exceeds a certain threshold is regarded as a detected source. In each
step we treat data of the GIS and the SIS separately. Finally, their
results are combined to improve statistics.

\subsection{Step~I: Source Detection}

In Step~I, we smooth the raw image (= the image in the sky coordinate
in the photon counts space) in order to effectively detect even faint
sources.  However, a simple convolution of the image data with an
axisymmetric, position-independent smoothing kernel would not be
sufficient, because the PSF depends strongly on the position of the
source in the focal plane of the XRT. Noticeably, the PSF elongates
toward the azimuthal direction and squeezed in the radial direction if
the point source is off the optical axis of the XRT (e.g., Takahashi
\etal\ 1995). We therefore smooth the images by cross correlating with
position-dependent PSFs in detector coordinates, which are converted
from the sky coordinates with information from the satellite attitude.  The
cross correlation function (CCF) is calculated as:
\[
	F(\vec{x}) = \int d\vec{x'} F_0(\vec{x'}) P(\vec{x}:\vec{x'}),
\]
where $\vec{x}$ represents positional coordinate, $F_0(\vec{x})$ is
the raw image, and $P(\vec{x}:\vec{x'})$ is the PSF intensity at
$\vec{x'}$ with peak at $ \vec{x}$. 
In calculating the CCF, proper modeling of the PSF is very important.
For the GIS, we adopted a synthetic PSF modeled as a two dimensional
Gaussian function squeezed in the radial direction as a function of the offset
angle (Ishisaki \etal\ 1995). For the SIS, we used the ray-tracing
program (Tsusaka \etal\ 1995) to reproduce the PSF at any position in
the focal plane.

Since the brightness profile of the background (CXB + NXB) peaks around
the optical axis and decreased to about 50\% at the edge of the GIS
FOV (Kubo \etal\ 1994; Ueda 1996), it becomes harder to detect sources
located at larger radii. Hence, we subtract the background from the
raw image before smoothing the GIS image. The background model is
produced by the superposition of 109 pointings onto blank skies whose
total exposure amounts to about 1 Msec. On the other hand, we neglect
this process for SIS data because the position dependence of the
background is much smaller due to its narrower FOV.

After smoothing the raw images separately for each sensor and
pointing, we superpose all the smoothed images into a single image
with a larger size.  Finally, we correct it for the exposure by
dividing with a superposed exposure map that are smoothed in the same
way. Figure~1 shows a contour map of the GIS image in the 0.7--7 keV
band thus created.

Next, we perform detection of source candidates from the smoothed (and
exposure-corrected) image created above. Since a point source appear
as a cluster of pixels having high counting rates in the smoothed
image, source candidates are picked up if the highest pixel in the
cluster exceeds some threshold level. The level is varied from a high
to a low level continuously and a newly appeared cluster is added to
the source-candidate list. This makes it possible to find all faint
sources completely even when brighter sources are located around them. 
The list of the identified clusters is used as inputs for the fitting
process in Step~II. The position of the highest pixel in each cluster
gives the source position.

\subsection{Step~II: Flux Calculation}

In Step~II we evaluate the statistical significances of the candidate
sources, and calculate their fluxes. The analysis is performed using
two-dimensional fitting to the raw image in the photon count space,
superposed from the multi pointings. An advantage of this method is
that we can determine the local background level, which is otherwise
difficult to estimate with enough accuracy to determine correct fluxes
of faint sources, from the observed image itself. We apply various
corrections, such as exposure and vignetting, to the model function,
not to the data.

The fitting model is basically a sum of the background image, and the PSFs
at the location of the source candidates. To reproduce the superposed sky
image in the model calculation, we sum up contributions of PSF and
background from different pointings weighted with exposures.  The
intensities of the sources and the background are free parameters to
be determined through the fitting. The source positions are fixed to
the values determined from the smoothed image in Step~I. Due to the
vignetting of the XRT, an observed count rate from a source with a
given flux decreases as the offset angle from the optical axis becomes
larger (Serlemitsos \etal\ 1995). To derive an intensity of a source,
we correct for this effect in the model calculation, so that we should
obtain the count rate when the source is observed at a reference
position in the focal plane. The reference position is taken to be a
8.5 arcmin (7 arcmin for the SIS) off axis position, and the count
rate is defined as that integrated within a radius of 6.0 mm
(corresponding to about 5.9 arcmin) for the GIS and 4.0 mm (about 3.9
arcmin) for the SIS. These vignetting correction factors are
calculated from the XRT response, which is well calibrated using
the Crab observations, as described in Appendix~A.
For the GIS, source peaks are modeled by the real PSFs obtained from
the Cyg X-1 observations (Takahashi \etal\ 1995; Ikebe \etal\ 1997),
and the background is from the same model, as used in Step~I
(superposition of real images of blank skies).
For the SIS,
the PSF and the CXB are reproduced by the ray-tracing program, and the
NXB is modeled using the night-Earth data (Ueda 1996). The systematic
errors in the PSF models are estimated to be 5\% and 15\% for the GIS
(Takahashi \etal\ 1995), and the SIS (Kunieda
\etal\ 1995), respectively, and are taken into account in the
fitting process. Finally, the transmission map of the grid support is
multiplied to the component of sources (PSFs) for the model of the
GIS.  Since the shape of the PSF and the vignetting correction factor
depend also on energy, we need to assume a spectrum of a source in the
model calculation. We adopted a power law with a photon index of 1.7
as a typical spectrum. The dependences of the PSF and the vignetting
correction factor on the assumed spectrum are negligible within a
reasonable region of the photon index (the difference of the vignetting
correction factor between $\Gamma = 1.0-2.0$ is less than a few percent).

Since the number of photons in one bin of the image is small, we use
Poisson maximum likelihood statistics to find the best fit parameters.
The maximum likelihood function is defined as:
$$
{\rm Likelihood} = \sum_{i,j} 
\{ 
-m(i,j) + d(i,j) \times \log m(i,j)
\}
,
$$
where $(i,j)$ represents the position of the bin, $m$ is the model, and $d$ is
the data. The sum is taken over the region used in the fitting. 
In this statistics, 1 $\sigma$ statistical error of one
parameter is defined as a deviation from the best fit value, when the
Likelihood decreased by 0.5 from its maximum.
The significance of detection is defined as
$({\rm best\ fit\ flux})/$
$({\rm 1\ \sigma\ statistical\ error\ of\ the\ flux})$.

For the fitting to converge correctly, the free parameters, namely
normalization of the background and those of peaks, are determined one by
one in the fitting process.  We first fit the image only with the
background component (CXB + NXB) to determine the background level.
Next, taking the background determined in this way, we determine the
flux of the brightest source. After the flux is obtained, this is then
frozen and the process is repeated for the next source. Once all the
detected source candidates are fitted, the parameters including the
normalization of background are allowed to float, producing final
parameters. Thus, any statistical coupling between the free
parameters are taken into account. In the procedure, we do not treat
all fields of the LSS simultaneously. Instead, we divided the entire image
into small regions ($19' \times 19'$ for the GIS and $14'\times14'$
for the SIS) so that the number of sources becomes appropriate ($<$29)
for one fitting process.  To take into account the PSF tails from
sources in outer region of the selected area, we fit the image of
larger size with some margins ($6'$ for the GIS, and $4'$ for the SIS)
when determining the flux of sources within the selected area.
We did not use data of a rectangular region surrounded
by four corners, $(RA(2000), DEC(2000))$ =
(13$^{\rm h}$19$^{\rm m}$5$^{\rm s}$, 33$^\circ 6'$), 
(13$^{\rm h}$20$^{\rm m}$15$^{\rm s}$, 32$^\circ 58'$), 
(13$^{\rm h}$21$^{\rm m}$3$^{\rm s}$, 33$^\circ 15'$), and
(13$^{\rm h}$19$^{\rm m}$52$^{\rm s}$, 33$^\circ 23'$), 
to avoid systematic errors due to a bright extended source located
there.

\subsection{The Source List}

First, we applied the methods described above to the combined image of
the LSS field separately for the GIS and the SIS in three energy
bands: the total band (0.7--7 keV), the hard band (2--10 keV; 2--7 keV
for SIS), and the soft band (0.7--2 keV). We thus obtained 6
preliminary source lists. At this stage we adopted relatively loose
detection criteria (above 3.2 $\sigma$) in selecting sources. We then
merged the three source lists in different bands to one list for each
detector. Due to statistical fluctuation, the detected positions of a
source can differ with energy bands. Here, we regarded any pair of
sources detected in different energy bands as identical, if their
positions coincide within 1.5 arcmin (1 arcmin for the SIS). These
criteria are reasonable, considering the position accuracy of each
detector. We adopted the position determined in the energy band in
which the maximum significance was obtained.  Finally, we combined the
GIS list and the SIS list, regarding sources detected with different
detectors within a distance of 1.5 arcmin as identical sources. We
adopted the positions determined by the SIS data, taking advantage of
its superior positional resolution (hence better position accuracy),
except for those located out of the combined SIS FOVs.  Because the
PSF of the XRT measured with the SIS has a sharp core, we can resolve
neighboring sources whose angular separation is as small as 1 arcmin.

Using the combined source list as inputs of source candidates, we
repeated Step~II in each energy band. Finally, we adopted the
following criteria for detection: (1) the significance of summed count
rate of the GIS and the SIS ($\sigma_{\rm t}$) should exceed 4.5, and
(2) the significance of either the GIS or the SIS ($\sigma_{\rm g}$ or
$\sigma_{\rm s}$) should also exceed 3.5. 
Studies based on Monte Carlo simulations justify the validity of setting
these criteria, as described below. Using $\sigma_{\rm t}$, the combination of
two significances measured with independent detectors, considerably
reduces contamination by fake sources (statistical fluctuation), since
real sources are expected to be commonly detected with both the GIS
and the SIS, while fake sources are not.

With these criteria, 
we detected 105, 44, and 72 sources, in the total, the hard, and the
soft energy bands, respectively. The total number of sources is 107. 
The number of sources detected both in two energy bands (or only in one
band) is summarized in Table~2. Table~3 gives the complete source list
in the LSS field with their positions, significances, and
vignetting-corrected count rates in the three energy bands. Note that
the GIS count rate of 1 c ksec$^{-1}$ in each energy band corresponds
to a flux of $6.4\times10^{-14}$ (0.7--7 keV), $1.1\times10^{-13}$
(2--10 keV), $4.5\times10^{-14}$ (0.7--2 keV)
\ergs\ , respectively, for a power-law spectrum with a photon index of 1.7. 
All the fluxes in this paper are corrected for the Galactic absorption
($N_{\rm H} = 1.1 \times 10^{20}$ cm$^{-2}$) to give unabsorbed
fluxes. In Table~4, we list the conversion factors from the count rate
to flux for several photon indices assumed.

\placetable{tbl-2}
\placetable{tbl-3}
\placetable{tbl-4}

\subsection{Spectra of the Detected Sources}

Spectral properties of these faint sources are of particular interest
in discussing the origin of the CXB. The count rates of the hard and the
soft bands already contain information of the spectrum for each source. To
derive more information, particularly spectra within the 2--10 keV
range, we repeated the image fitting procedure (Step~II) in finer
energy bands (2--4 keV and 4--10 keV) for the GIS data. Then, we produced
spectra for each source in two energy bands for the SIS (0.7--2 and
2--7 keV), and in three energy bands (0.7--2, 2--4, and 4--10 keV) for
the GIS. For each source, we fitted the GIS and the SIS spectra
simultaneously, assuming a power law with an absorption fixed at the
Galactic column density of $N_{\rm H}$ = $1.1\times10^{20}$
cm$^{-2}$.  In the spectral fitting, we used an Auxiliary Response Function
(ARF), constructed for a point source located at the reference position 
defined above, where the obtained count rate is normalized.
The best-fit power law indices obtained are given in a column of
Table~3, with 1$\sigma$ statistical errors. Figure~2 shows a
correlation between a flux and a photon index determined in the
0.7--10 keV range. In the figure, the flux (0.7--7 keV) is 
converted from the GIS count rate in the 0.7--7 keV band, assuming a
photon index of 1.6 (an average index for a faint source sample; see below).

In order to see the statistical properties of these sources, we next
made average spectra of sources for the three survey bands (the total,
hard, and soft band) by summing up the spectrum of individual sources. 
Since this operation corresponds to calculating a flux-weighted
average, very bright sources dominate the signal and significantly
affect the result.  To study the property of faint sources, we
therefore use flux-limited samples for each survey band: sources with
fluxes (in units of \ergs) of $0.5 - 2.0\times10^{-13}$ (0.7--7 keV)
detected in the total band, those with fluxes of $0.8 -
4.0\times10^{-13}$ (2--10 keV) detected in the hard band, and those
with fluxes of $0.2 - 2.3\times10^{-13}$ (0.7--2 keV) detected in the
soft band.  We hereafter refer to them as the total-band sample, the
hard-band sample, and the soft-band sample, each of which consists of
74, 36, and 64 sources, respectively.

We fitted these average spectra with a power law with the Galactic
absorption. Since the SIS spectra do not cover the energy band above 7
keV, we here use only the GIS spectrum to make the statistical weight
uniform over the entire 0.7--10 keV range. The spectral fit is
performed in two energy ranges: 0.7--10 keV band consisting of three
bins (0.7--2, 2--4, and 4--10 keV) and 2--10 keV band consisting of
two bins (2--4 and 4--10 keV). The fitting results are summarized in
Table~5. We obtained an average photon index of $1.63\pm0.07$
(1$\sigma$ statistical error) in the 0.7--10 keV range for the
total-band sample. The 2--10 keV photon indices are 1.63$\pm$0.18,
1.49$\pm$0.10, and 1.85$\pm$0.22 for the total, the hard, and the soft
band sample, respectively (the 2--10 keV index of the hard band sample
is not exactly same as the value reported in Ueda \etal\ (1998)
because of inclusion of $\sigma_{\rm s}$ in the detection criteria).

It should be noted that the average spectrum in the 0.7--10 keV band
for the hard-band sample (or the soft-band sample) is subject to a
statistical bias in the sense that average spectrum of the hard-
(soft-) band sample tends to show a harder (softer) spectrum than in
reality: if sources are selected based on significances only in the
hard (soft) band, there should be a sampling effect that a sample
selectively contains sources whose ``observed'' spectra in the 0.7--10
keV band are hard (soft) merely due to statistical fluctuation. Hence,
the two values in parenthesis in Table~5 should not be considered to
reflect a real property, without correcting for these biases. On the
other hand, no such bias exists, when we discuss a spectral property
in the 2--10 keV range for all the samples.

\placetable{tbl-5}

\section{Verification}

\subsection{Flux Uncertainty}

Here we evaluate systematic errors in the obtained fluxes and spectra.
In general, systematic errors can be divided into two categories. In
the first category are errors caused by the calibration uncertainty. 
Appendix~A evaluates this error in detail. The uncertainty in the
absolute photometry of \asca\ is conservatively estimated at 10\%. 
The error in the relative flux is estimated to be 8\% for an
individual source, and is much smaller for an average of many sources.
Note that this error is smaller than the statistical error for a faint
source detected with a significance less than 12$\sigma$
(1/12 $\sim$ 8\%). In the second category are errors
associated with the improper analysis method or those inevitably
caused by principles (such as the source confusion), which remain,
even if we fully understand the response functions of instruments.

To begin with, we compared the fluxes obtained from the GIS and the
SIS to check the consistency within our data.  We have 71 sources
detected with both the GIS and the SIS above 3.5$\sigma$ (and
$\sigma_{\rm t} > 4.5$) in the total band. Figure~3 shows a comparison
in the form of a scatter-plot between the fluxes obtained with the
SIS and the GIS for this sample. Since the spectrum of an individual
source at faint flux levels includes a large statistical error, we
assumed a photon index of 1.6 for all sources to convert a count rate
to a flux. As shown in the figure, we found a good correlation between
the GIS and the SIS fluxes.

Studies by Monte Carlo simulation are indispensable to check
systematic errors of the second category, which are otherwise
difficult to evaluate. We developed an instrument simulator that
simulates the response of \asca\ (Hirayama \etal\ in preparation; an
example of analysis utilizing it is given in Ikebe \etal\ 1997). The
complicated responses of the instruments, in terms of PSF, vignetting,
grid shadow of the GIS, etc, are taken into account in the simulator. 
The simulator produces a photon-event list from the input sky image and
spectrum, according to the probability distribution in the responses,
which are well calibrated (Appendix~A).

To simulate the LSS data, we generated photon-event lists with the same
conditions (such as exposure and satellite attitude) as the LSS
observations. Sky images were created according to the \logn\ obtained
by Ueda \etal\ (1998), assuming that it extends towards a lower flux
level with a function of $N(>S) \propto S^{-3/2}$, until the whole
intensity of the CXB is explained by integral of sources. For
simplicity, all the sources are assumed to have a single spectrum
regardless of the flux level: a power law with a photon index of 1.7
over the full energy range (0.7--10 keV).  The NXB was added based on
the night Earth data, and we fluctuated its intensity from pointing to
pointing by a factor of 10\%, considering the uncertainties 
in the modeling of the NXB (Ishisaki \etal\ 1997). We then analyzed the
simulated data obtained in this way, and applied our source finding
procedure in exactly the same manner as for the real data. To reduce
the statistical fluctuation in the simulated result, we ran the
simulations several times, with a corresponding exposure of 23 Msec
in total.

Figure~4(a) shows a comparison of the calculated flux with the input
flux for these simulated data. Here, we define the input flux as the
flux of the brightest source within a 1 arcmin radius around the
detected position, considering the angular resolution of the PSF. As
noticed from the figure, we obtain a good agreement between input and
output flux in general, although there is a tendency for the observed
fluxes to become larger than the input ones, when the flux is below
$\sim 2\times10^{-13}$ \ergs\ (0.7--7 keV). This can be attributed to
two effects. One is the bias due to the statistical fluctuation of
photon numbers: we selectively detect sources whose output fluxes are
larger than the input fluxes around the sensitivity limit, because the
significance is determined by an output flux (i.e., observed count
rate) rather than by an input flux. The second effect is the source
confusion by nearby sources: the observed flux contains all the
integrated flux from sources that cannot be resolved around the
detected position.

Figure~4(b) shows a histogram of the ratio of the input flux to the
output flux. In the same figure, we also show a similar histogram,
marked with a dashed line, for the case when we define the input flux
as an integrated (not the brightest) flux within a radius of 1 arcmin
around the detected position. As expected, the ratio (input/output)
becomes systematically larger for the latter definition. The
difference of the mean value between these two histograms
quantitatively gives the effect of the source confusion. We thus
estimate the contamination by nearby sources in the observed flux to
be about 10\% on average. This means that effects due to the source
confusion are not serious in our analysis, since the number density of
the detected sources, derived below ($\sim$ 10--20 deg$^{-2}$), is
small enough compared with the confusion limit determined by the beam
size ($\sim$ 1 arcmin).

From Figure~4, it can be seen that there are several sources detected
with an extremely high flux, relative to the input flux. These are
considered to be not real sources but ``fake'' sources, produced by the
statistical fluctuation of the background. We estimate the fraction of
fake sources to be at most a few percent in the overall detections. In
other words, the probability to pick up a fake source by our source
finding method is about $\sim$ 0.5 deg$^{-2}$ with the detection
criteria we adopted in this paper (4.5$\sigma$).

We also check the accuracy of the output spectra. From a scatter plot
between a photon index and a flux obtained for the simulated data, we
confirmed that the output photon index scatter around the input value
(1.7) and there is no significant systematic change with respect to
the flux level. We also verified that the summed spectra for the
flux-limited samples, defined above, are consistent with the input
spectrum within a statistical error. These results confirm the
validity of the spectrum determination by our analysis method.

\subsection{Position Accuracy}

The source position in the source list contains (i) an error arising
from the uncertainty of the absolute attitude solution, (ii) a
statistical error due to limited number of photons, and (iii) a
systematic error caused by source confusion. As described earlier,
the first term is evaluated to be 0.5 arcmin in radius (90\%
confidence level). To evaluate the combined error of the remaining two, we
again analyze the simulated data. Figure~4(c) shows a histogram of
angular distance between the detected position, and the position of the
brightest source within a 1 arcmin radius, which should be taken as a
real counterpart. According to this histogram, the error comes out to
be 0.6 arcmin (90\% confidence level). When limiting for sources that
are detected also in the hard (or the soft) band with significances
above 4.5$\sigma$, we found the error reduces to 0.4 arcmin. Hence,
taking a root sum square of all possible errors, we estimate the final
positional error to be 0.6--0.8 arcmin in radius (90\%), depending on
the detection significance. To improve it further, observations with
the \rosat\ HRI are ongoing.

\section{The \Logn}
\subsection{Relation of Observed Area and Sensitivity}

The \logn\ (number-flux relation; $N(S)$ in the differential form and
$N(>S)$ in the integral form) is the most fundamental measure which
directly describes the contribution of individual sources to the CXB.
To derive the \logn, it is necessary to calculate the survey area,
$\Omega$, as a function of flux $S$, or in other words, to determine
the sensitivity limits over the entire survey field. The sensitivity
limits depend on parameters such as exposure, effective area
(vignetting), the background level, and the shape of the PSF. It is
not an easy task, however, to compute it by an analytic formula due to
the following reasons: (1) the image is composed of multiple
pointings, and the above parameters are different from position to
position. (2) Our source-finding method is based on image fitting, and
thus the estimates of the sensitivity limits are not trivial.

To estimate the detection significance for a given flux at any
position in the survey field, we utilized the model function used in
the source finding procedure. We first compute an image model around
the source position, which consists of a point source with the given
flux and the background. Then we fit the computed image with the same
model function, varying the flux of the source as a free parameter. 
Using the same principle in the fitting process given in Section~3.2,
we can derive the statistical error of the flux, thus the significance
of detection, expected for the given flux. We confirmed the validity
of this method by comparing the result with a more realistic
simulation, in which the number of photons fluctuates according to
Poissonian statistics. This procedure is repeated for several different
fluxes, as to determine the sensitivity limits over the entire field for
given detection criteria, which thus enables us to compute $\Omega(S)$.  In
practice, we define $S$ as a count rate (corrected for the vignetting)
rather than as a flux, because the detection significance is simply
extracted from a count rate.

We compute $\Omega(S)$ separately for the GIS and the SIS, because the
ratio of count rates between the GIS and the SIS depends on the source
spectrum, which makes it difficult to combine the two results.  Thus,
when deriving $\Omega(S)$, and hence the \logn, we adopted the
following detection criteria applied separately to the GIS or the SIS:
(1) $\sigma_{\rm g}$ $>3.5$ ($\sigma_{\rm s}$ $>3.5$ for the SIS), and
(2) $\sigma_{\rm t}$ $>3.5$. Note that the criteria are different from
those adopted in making the source list (Section~3.3). The second
condition is added just to reduce the fraction of fake sources.
Figure~5(a) shows results for $\Omega(S)$ computed for the three
survey bands. The GIS achieves higher sensitivity than the SIS in
the hard band, while the SIS achieves higher sensitivities in the
total and the soft bands, as is expected from the energy dependence of
their detection efficiencies. The number of sources detected with the
same detection criteria per unit logarithmic flux range are plotted in
Figure~5(b). Dividing these numbers by $\Omega(S)$, we can derive
``observed'' \logn, $N(S)$, in the differential form. In plotting
Figure~5, we coverted the count rate to flux assuming a photon index
of 1.6 (0.7--7 keV), 1.5 (2--10 keV), and 1.9 (0.7--2 keV).

\subsection{Biases on the Observed \Logn s}

The ``observed'' \logn s are, however, considered to be subject to
some biases.  They include (1) bias induced by the statistical
fluctuation of the measured flux (Eddington's bias), (2) effect of the
source confusion, (3) inefficiency of the source detection, and (4)
contamination by fake sources. Again, Monte Carlo simulations are the
most powerful mean to estimate these biases. Using the set of
simulated data created in Section~4, we calculated $\Omega(S)$ and
derived the \logn\ in the same manner as for the real data.

Figure~6 show the results of integral \logn s derived from the
simulated data. The faintest points in the total and the soft band are
determined mainly by the SIS, while those in the hard band are
determined by the GIS. The
input \logn\ is plotted by a dashed line in the same figure.  As
noticed, the results of the simulation are almost consistent with the
input parameters, although week systematic deviations can be found. 
The points near the sensitivity limit in the total and soft band
determined by the SIS data are located slightly above the input line by
10--20\%. In the soft band, the GIS result are also
$\sim$10\% above the input at the flux level below $\sim 10^{-13}$
\ergs\ (0.7--2 keV). The reason why this deviation is seen only in the
soft band can be explained in terms of the source confusion, since the
positional resolution of the GIS becomes worse for lower energy
(Ohashi \etal\ 1996) and the effect should be most significant in the
soft energy band.

\subsection{The Integral \Logn s}

Finally, we derived \logn s from the real data correcting for these
biases. From the simulation, we found an empirical formula can be
applied to correct them, by scaling the ``observed'' integral \logn,
$N(>S)$, by a flux-dependent factor, $f$, which is defined as:
{\scriptsize
\begin{eqnarray*}
({\rm For\;} N>N_0) \;\;\;\; f &=& 1 / [0.1 \times ({\rm log}N-{\rm log}N_0) / ({\rm log}N_1-{\rm log}N_0) + 1]\\
({\rm For\;} N<N_0) \;\;\;\; f &=& 1,
\end{eqnarray*}
}
where $N$ is the observed integral source counts in units of
deg$^{-2}$.  The scaling factor becomes 10\% at the source number
density of $N$ = $N_1$, and 0\% at $N < N_0$.  We set $N_0 =
15$~deg$^{-2}$ and $N_1 = 30$~deg$^{-2}$ for the SIS results for any
band, and $N_0 = 3$~deg$^{-2}$ and $N_1 = 15$~deg$^{-2}$ for the GIS
results only in the soft band. No correction is applied for the GIS
results in the other bands since no significant biases were seen in
the simulation. Note that this formula can be applied only in the
currently interesting number-density range ($N^<_{\sim}N_1$).

Figure~7 shows the final results of integral \logn s in the three
survey bands after applying this formula. Here, we convert the count
rate in each band into the flux in the same energy band, assuming a
photon index of 1.6 (0.7--7 keV band), 1.5 (2--10 keV band), and 1.9
(0.7--2 keV), each of which corresponds to a photon index of the
average spectrum of the faint-source sample, as defined in
Section~3.4.  The 90\% statistical errors in the source number counts
are plotted at several data points.

\section{Discussion}

\subsection{The \Logn}

From the LSS, we determined the \logn\ by direct source counts in the
three survey bands, 0.7--7 keV, 2--10 keV, and 0.7--2 keV.  These
results were obtained by the careful analysis based on extensive
instrumental calibration and detailed simulation studies.  In the
0.7--7 keV band, where the sensitivity is optimized for most of
sources except for very soft or hard sources, we found that the \logn\
is well represented by $N(>S)$ $\propto$ $S^{-3/2}$ over a wide flux
range from $7\times 10^{-14}$ to $8 \times 10^{-13}$ \ergs\ (0.7--7
keV). As already reported in Ueda
\etal\ (1998), the 2--10 keV \logn\ lies on the extrapolation from the
previous results by \heao\ A2 (Piccinotti
\etal\ 1982) and \ginga\ (Kondo \etal\ 1992) with a euclidean slope of
$-3/2$. The result is consistent with the fluctuation analysis by
ASCA (Gendreau, Barcons, \& Fabian 1998).

To compare our results in the 0.7--2 keV band with previous studies,
we superpose the \logn\ obtained by \rosat\ (Hasinger 1998) on our
results in Figure~7 (c).  Here we convert the flux from 0.5--2 keV to
0.7--2 keV assuming a photon index of 1.9. As shown in the figure, our
results are consistent with the \rosat\ results (and also with
\einstein\ results; see Primini \etal\ 1991) within statistical error,
although we find a discrepancy that the
\asca\ source counts is slightly higher by $20\sim30\%$.

There could be two explanations for this discrepancy. One is due to
the calibration uncertainty in the flux (Appendix~A). In addition to
the uncertainty of the absolute photometry ($\sim$10\%), the low
energy response of the \asca\ has an uncertainty of $\sim
3\times10^{20}$ cm$^{-2}$ in a neutral hydrogen column density, which
could produce an error of $\sim$5\% in the 0.7--2 keV flux. Note that
a systematic error of 10\% in the flux leads to $\sim$15\% apparent
difference in the source counts.  The other explanation could be
attributed to the nature of the detected sources.  Even though \asca\ and
\rosat\ cover the common energy band of 0.7--2 keV, the ``effective''
band at which detection efficiency is maximized is different: the
\asca\ GIS/SIS has better detection efficiency toward higher energy in
the 0.7--2 keV range (e.g., Ohashi
\etal\ 1996), while the \rosat\ PSPC has a peak efficiency around 1
keV (Pfeffermann \etal\ 1986) and also covers the energy range below
0.7 keV. If some fraction of sources has hard energy spectra in the
0.7--2 keV range (e.g., absorbed spectra), they are more easily
detected with \asca\ than \rosat. For example, if half sources in the
soft-band sample have absorbed spectra with $N_{\rm H} = 3 \times
10^{21}$ cm$^{-2}$ with a photon index of 2.2 and the other sources
are unabsorbed, discrepancy of $\sim$10\% in the source counts can
be produced, if we simply take a photon index of 1.9 
in the flux conversion between \rosat\ and \asca\ assuming a single
population.

\subsection{The Resolved Fraction of the CXB}

The resolved fraction of the CXB depends on the absolute intensity of
the CXB emission, which can be different by about 20\% in the
literature (e.g., Marshall \etal\ 1980; Gendreau \etal\ 1995; Ishisaki
\etal\ 1999). To eliminate uncertainty in the absolute photometry
of instruments and the field-to-field fluctuation of the surface
brightness of the CXB, we adopt the result by Ishisaki \etal\ (1999)
that are obtained with the same instrument (GIS) from the same field
(LSS) as our data. Correcting for the contribution of brighter sources
than the brightest sources in the LSS field, we use $(7.0\pm0.7)\times
10^{-8}$ \ergss\ (0.7--7 keV) and $(6.5\pm0.7)\times 10^{-8}$
\ergss\ (2--10 keV) as the whole CXB intensity. Note that this 2--10 keV flux 
is larger by 23\% than the result by Marshall \etal\ (1980). 
The error (10\%) shows the systematic error due to
uncertainty of the evaluation of the stray light in the analysis of
the CXB spectrum (Ishisaki \etal\ 1999). The
\logn s obtained in this paper indicate that $30(\pm3)$\% of the CXB has
been directly resolved into discrete sources with a sensitivity limit
of $6\times10^{-14}$ \ergs\ in the 0.7--7 keV band, and $23(\pm3)$\%
with a sensitivity limit of $1\times10^{-13}$ \ergs\ in the 2--10 keV
band. The attached error merely comes from the uncertainty of the CXB
intensity described above.

\subsection{Comparison of \Logn s in the Soft and the Hard Band}

The comparison of \logn s between soft and hard bands provides
spectral information over the whole 0.7--10 keV energy band. 
Assuming a single population of sources, we can constrain the
average spectrum of sources so that the hard and soft \logn s should
be consistent. Previously, this kind of comparison has been made, based
on results of different satellites (such as \ginga\ and \einstein),
which could be subject to uncertainty in the mutual calibration
between different instruments. Since \asca\ covers the energy bands
above and below 2 keV with single detectors, we can now discuss it
with much less uncertainties than in previous studies. Figure~8 shows the
\logn\ in the soft band (0.7--2 keV), superposed on the \logn\ in the hard
band (2--10 keV). The 0.7--2 keV count rate is converted to the 2--10
keV flux, assuming two power-law photon indices (1.4 and 1.7). As
noticed from the figure, the two \logn s coincide with each other if
we choose a photon index 1.5--1.7 at a flux level of $(1\sim2) \times
10^{-13}$ \ergs\ (2--10 keV). Indeed, the result is consistent with
the average spectrum of the total-band sample, which shows a photon
index of 1.63$\pm$0.07 in the 0.7--10 keV range.

\subsection{Spectral Properties of the Detected Sources}

The most important result is finding evidence for the solution of the
spectral paradox (Ueda \etal\ 1998). The average spectrum of the hard
band sample, consisting of sources detected in the hard band with
fluxes of $(0.8-4)\times 10^{-13}$ \ergs\ (2--10 keV), shows a photon
index of $1.49\pm0.10$ in the 2--10 keV range (1$\sigma$ statistical
error in the mean value). This is significantly harder than the
spectrum of sources at much brighter flux level obtained by previous
studies, such as the fluctuation analysis with \ginga\ that showed a
photon index of $1.8\pm0.1$ (Butcher \etal\ 1997). Moreover,
the photon index is very close to that of the CXB in the 2--10 keV
range, $\Gamma \simeq 1.4$ (e.g., Gendreau \etal\ 1995; Ishisaki \etal\
1999), to within statistical errors. 

When the energy range is expanded to below 2 keV, the photon index in
the 0.7--10 keV range for the total band sample, $\sim$1.6, also
suggests a similar hardening of average spectrum of sources with
deceasing flux. The spectrum of the CXB can be approximated with a
power law of $\Gamma \simeq 1.5$ in the 0.7--10 keV range when the soft
component below 1.2 keV is appropriately modeled (Gendreau \etal\
1995; Ishisaki \etal\ 1999). Thus, even in the 0.7--10 keV range, the
average spectrum of sources becomes close to the CXB spectrum.

These spectra of sources at a flux level of $\sim 10^{-13}$ \ergs\
(2--10 keV) are harder than a typical spectrum of bright type-I AGN,
both in the 2--10 keV and 0.7--10 keV range (e.g., Williams \etal\
1992; Nandra \& Pounds 1994; George \etal\ 1998). This implies that a
population of sources with hard energy spectra that is responsible to
the CXB emission begins to dominate in the sensitivity limits we
achieved.

\subsection{Populations in Our Sample}

Using results from optical identifications in the \rosat\ surveys, we
can make a rough estimate of the fraction of well-known X-ray
populations in our survey, such as type-I AGNs, clusters of galaxies
(CG), and stars. In the 0.7--2 keV band, their fractions are expected
to be same as the \rosat\ survey. According to the results by
\rosat\ at a flux of $10^{-13}-10^{-14}$ 
\ergs\ (0.5--2 keV) (e.g., Hasinger 1996; Bower \etal\ 1996), 
about 60\% of the total sources are identified as type-I AGNs and 10\%
are clusters of galaxies. The fraction of Galactic stars depends both on
the flux limit and on the Galactic latitude. We estimate that 10--15\% of the
LSS sources detected in the soft band can be stars, considering the
high Galactic latitude of the LSS field.

The other populations contain objects identified as narrow emission
line galaxies (NELGs), and unidentified objects in the \rosat\ surveys. 
Recently, however, Hasinger \etal\ (1998) and Schmidt \etal\ (1998)
pointed out that a part of the previous identifications with NELG
using the Position Sensitive Proportional Counter (PSPC) could be
incorrect because of a large position uncertainty and the source
confusion problem. Indeed, the ultra deep survey with the \rosat\ HRI
shows 60--70\% of the detected sources are type-I AGNs,
contradictory to the previous prediction. According to those authors, 10--20\%
of the total sources could be mis-identified in the deep surveys by
\rosat\ on which we base our estimation. To take this possible
mis-identification into account, we have assumed the extreme case, that
20\% of the total sources that belonged to other population are
actually type-I AGNs. Estimates based on this assumption are presented
in parenthesis in columns of Table~6.

For the sample of the 0.7--7 keV and 2--10 keV survey, we can estimate
the number density of each population, thus its fraction, by
converting the sensitivity limit in each survey band to that in the
soft band with assumption of spectra over the 0.7--10 keV range. Note
that because the sensitivity limit is determined by the count rate,
the flux limit in units of \ergs\ depends on spectra (Table~4), and
hence can differ with populations. As for the absolute source counts in
the soft band, we adopt our results derived from the LSS to eliminate
any possible uncertainty in the mutual calibration between \rosat\ and
\asca.

This estimate is subject to the assumed spectral shape. For the
spectrum of type-I AGNs, we assume two cases: a power law with a
photon index of 2.0 and 1.6 in the 0.7--10 keV range. Taking these
spectra as two extreme limits seems to be reasonable, according to
observations of type-I AGNs with \asca\ at brighter flux levels
(George \etal\ 1998; Cappi
\etal\ 1997). For clusters of galaxies, we assume a Raymond-Smith
plasma model with a temperature of 6 keV, located at a redshift of 0.2,
which are typical parameters of clusters observed with \asca\ (Mushotzky and
Scharf 1997). For stars, we assume a power law with a photon index of
$3.0$ over the 0.7--10 keV range. Since contribution of clusters of
galaxies and stars are relatively small, our estimate is not much
affected by these assumptions.

The results of our estimates are summarized in Table~6. The fraction
of type-I AGNs is estimated to be 45--75\% in the 0.7--7 keV survey,
and 30--70\% in the 2--10 keV survey. Then, the fraction of other
population of objects (other than type-I AGN, clusters, or stars)
becomes 5--40\% in the 0.7--7 keV survey, and 20--60\% in the 2--10
keV survey bands, respectively.  The corresponding number densities
are $N(>S)$ = 1.5--12 deg$^{-2}$ at a flux limit of $7\times10^{-14}$
\ergs\ (0.7--7 keV) and $N(>S)$ = 3--9 deg$^{-2}$ at $1\times10^{-13}$
\ergs\ (2--10 keV). In order to reproduce 
the observed average spectrum of the LSS sample, which has a photon
index of 1.63$\pm$0.07 in the 0.7--10 keV range and 1.49$\pm$0.10 in
the 2--10 keV range, these other population of sources must have hard
spectra.  We found that the photon index of this population is
expected to be 1.1--1.5 in the 0.7--10 keV range and 0.9--1.2 in the
2--10 keV range within the uncertainty range given in Table~6.

\placetable{tbl-6}

We can expect that most of these hard sources are detectable by
\rosat\ with sufficient exposures, knowing their 
average spectrum calculated above ($\Gamma =$ 1.1--1.5 in the 0.7--10
keV range). To check the consistency of the discussions, we
examine here whether the expected number density of these sources can be
explained by that of the ``other'' populations (other than type-I AGNs,
clusters, or stars) detected in the \rosat\ surveys. If we assume a
photon index of 1.1 over the 0.5--10 keV range, our sensitivity limit
in the 2--10 keV band, $1\times10^{-13}$ \ergs, corresponds to
$2\times10^{-14}$ \ergs\ in the 0.5--2 keV band, which is fainter than
the flux limit of the LSS soft-band survey, but brighter than that of
the deep surveys with \rosat\ (Hasinger 1996 and references therein). 
According to the \rosat\ results, the number density of the ``other''
populations is estimated to be $\sim$8 deg$^{-2}$ at this flux limit, if
their number fraction in the total source counts is 20\%. This
number density is consistent with that of the hard sources in our
survey, 3--9 deg$^{-2}$ at a flux limit of $1\times10^{-13}$
\ergs\ (2--10 keV). This implies that at least a part of \rosat\
sources that belong to the ``other'' populations have hard energy
spectra over the 0.5--10 keV range.

Good candidates of this population with hard spectra are type-II AGNs
(Antonucci and Miller 1985; Awaki \etal\ 1991). Indeed, the hardest
source in the LSS was optically identified as a type-II Seyfert galaxy
(Akiyama \etal\ 1998), with a heavily absorbed X-ray spectrum (Sakano
\etal\ 1998). Preliminary results of on-going optical identification
indicate that 20--30\% of the hard-band sample are type-II AGNs
(Akiyama \etal\ in preparation), which is consistent with the above
estimate. From the optical identifications, we find a good
correlation that the hard sources correspond to type-II AGNs, which
strongly supports our interpretation.

A comparison of our results with the prediction from the population
synthesis model proposed by Madau \etal\ (1994) and Comastri \etal\
(1995), in which type-II AGNs play an important role to reproduce the
CXB, would be interesting. According to Comastri \etal\ (1995), the
fraction of type-II AGNs in the total source counts is about 40--50\%
at a flux of $10^{-13}$ \ergs\ (2--10 keV). To connect this prediction
with our results, however, we must correct for the difference of flux
limits for type-I and type-II AGNs that have different spectra.  Just
to see this effect, we assume a photon index of 1.7 for type-I AGN and
1.0 for type-II AGN in the 2--10 keV range. Then the flux limit for
type-II AGNs is brighter than type-I AGNs by about 20\% (Table~4),
which corresponds to a factor of 30\% in the integral source counts
assuming a slope of $-3/2$. Thus, the fraction of type-II AGNs in the
LSS hard-band sample is predicted to be 30--40\%. At present, this is
consistent with our estimate, and with the optical identifications,
considering uncertainty in the several assumptions we have made. A
more detailed study using the optical results will be presented
elsewhere.

\section{Conclusion}

We presented a complete source list, containing 107 sources from the
three survey bands of 0.7--7 keV, 2--10 keV, and 0.7--2 keV, in the
LSS field. Based on extensive calibration and simulation study, the
results are carefully verified. We derived the \logn s in each survey
band with a flux range of $6\times 10^{-14}$ to $8\times10^{-13}$
\ergs\ (0.7--7 keV), $1\times 10^{-13}$ to $5\times10^{-13}$ \ergs\
(2--10 keV), and $3\times 10^{-14}$ to $2\times10^{-13}$ \ergs\
(0.7--2 keV). At these flux limits, 30($\pm3)$\% of the CXB in the
0.7--7 keV and 23($\pm3$)\% in the 2--10 keV band have been directly
resolved into discrete sources. We found that faint sources around the
sensitivity limits show on average a photon index of 1.5 in the 2--10
keV range, and 1.6 in the 0.7--10 keV range. These spectra are harder
than a typical spectrum of type-I AGNs, suggesting that contribution
of a population with hard energy spectra becomes significant at a flux
level of $\sim 10^{-13}$
\ergs\ (2--10 keV). 
With reasonable assumptions, the fraction of these hard sources is
estimated to be 20--60\% in the hard band sample. Type-II AGNs are
good candidates for this population of sources, which is being confirmed by
the on-going optical identifications.

\acknowledgments

We thank members of the LSS team and the ASCA team for their support
in planning observations, satellite operation, and data acquisition. 
We acknowledge the ASCA\_ANL and SimASCA software development teams
for supporting the analysis technique.  M. S. and M. A. acknowledge
the supports from the Japan Society for the Promotion of Science for
Young Scientists. We are also grateful to M. Freund for his
careful review of the manuscript.

\appendix

\section{Calibration of Instrument Response}

Calibration of instrument response, including that of the absolute
photometry, is crucial in our study, particularly when we make
comparisons with other missions. Here we describe the calibration
method and evaluate systematic errors in the instrument response used
in the paper.

\subsection{GIS+XRT}

The response of the GIS and the XRT has been finally determined by the
observations of Crab nebula, which is a standard calibration source in
X-ray astronomy. Fukazawa, Ishida, \& Ebisawa (1997) summarize the
results of spectral fits for the Crab data taken at various positions,
using the released GIS+XRT responses, which have been tuned
especially at the nominal pointing position. They found a tendency
that the observed absolute flux increases as the radius from the
optical axis becomes large. Although there is no such tendency in the
photon index and the column density, the average column density is
\nh\ $\sim 2.5\times10^{21}$ cm$^{-2}$, which seems to be somewhat
lower than a standard value ($(2.7-3.3)\times10^{21}$ cm$^{-2}$, Toor
and Seward 1974).

To improve the accuracy of the response further, we corrected it in two
practical ways. First, we scaled the effective area of the XRT
by an energy-independent factor given as a function of the radius from the
optical axis, with a similar formula described in Ishisaki (1996). 
Second, we modified the low energy efficiency so that a larger column
density should be obtained by \nh\ = $3\times10^{20}$ cm$^{-2}$.  Note that
this operation is not contradictory to previous calibrations,
considering the uncertainty in the low energy response (Ueda 1996).

The fluxes from Crab obtained with the modified response, used
throughout in this paper, are plotted in Figure~9 against the radius
from the optical axis. As noticed from the figure, the scatter of the flux
is less than 8\%, which corresponds to the amount of systematic error
in relative fluxes observed at different pointings. The average
parameters obtained are flux = $2.16\times10^{-8}$ \ergs\ (2--10 keV),
photon index = 2.08, and \nh\ = $2.8\times10^{21}$ cm$^{-2}$, all of
which are consistent with the standard values (Toor and Seward 1974). 
However, we conservatively attach 10\% error in the absolute flux,
considering the scatter of the flux between different satellites
(Seward 1992). Similarly, the systematic error in \nh\
is estimated to be $3\times10^{20}$ cm$^{-2}$, which produces
additional 5\% uncertainty in the 0.7--2 keV flux.

\subsection{SIS}

The degradation of detection efficiency of the SIS, which is prominent
in 4 CCD mode, becomes a problem in our study (Dotani \etal\ 1995).
This effect is energy independent to first order. To calibrate it, we
examined the change of apparent count rate of the CXB spectra using
the LSS data taken in different periods, separated by 6 months over 2
years. Since the LSS covers a wide area even in one period, the
fluctuation of the CXB surface brightness is almost canceled out after we
discard fields where extreme bright sources are located. As expected,
we found a clear decrease of the count rate with time. Also,
a comparison with the GIS data suggests that even at December 1993, only
10 months after launch, the absolute detection efficiency in the 4 CCD
mode degraded to about 80\% relative to that in the 1 CCD mode at 
launch. We correct this with formula of
\begin{eqnarray*}
f&=&0.0033 \times t + 0.083\;\;\; ({\rm SIS0})\\
 &=&0.0043 \times t + 0.084\;\;\; ({\rm SIS1}),
\end{eqnarray*}
where $f$ is detection efficiency normalized by that in the 1 CCD mode
at launch, and $t$ is the time in months since the launch. Note that this
formula can be applied only for a period of $t=10-28$. The good
correlation between the GIS and SIS fluxes in Figure~3 demonstrates
the validity of this correction within accuracy of 5\%.


\onecolumn

\appendix
\setcounter{figure}{0}

\begin{figure}
\epsscale{1.0}
\plotone{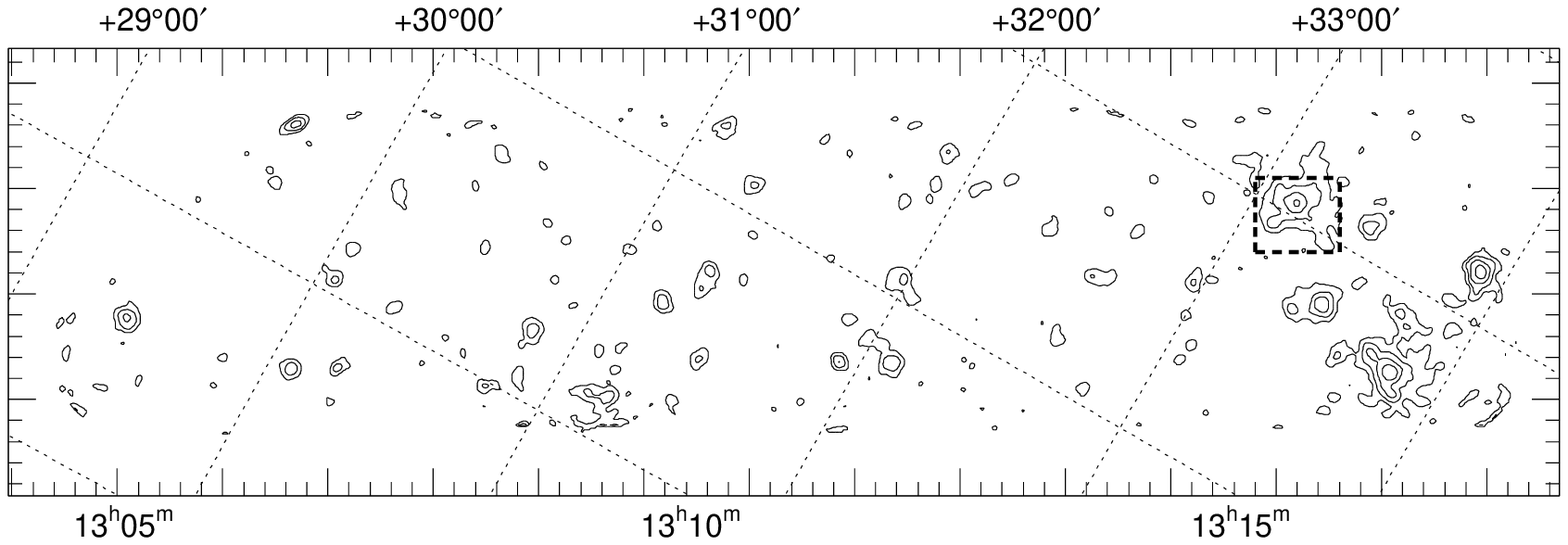}
\caption{
The contour plot of the GIS smoothed images of the LSS field in the
0.7--7 keV band obtained by cross-correlating with the
position-dependent PSF. The exposure is corrected for each point by
dividing with the smoothed exposure map.  The rectangle surrounded by
dashed lines represents a region where our analysis (Step~II; see
text) was not performed, to avoid systematic errors due to a bright
extended source located there.
\label{fig1}}
\end{figure}

\placefigure{fig1}

\begin{figure}
\epsscale{0.8}
\vspace{0.5cm}
\plotone{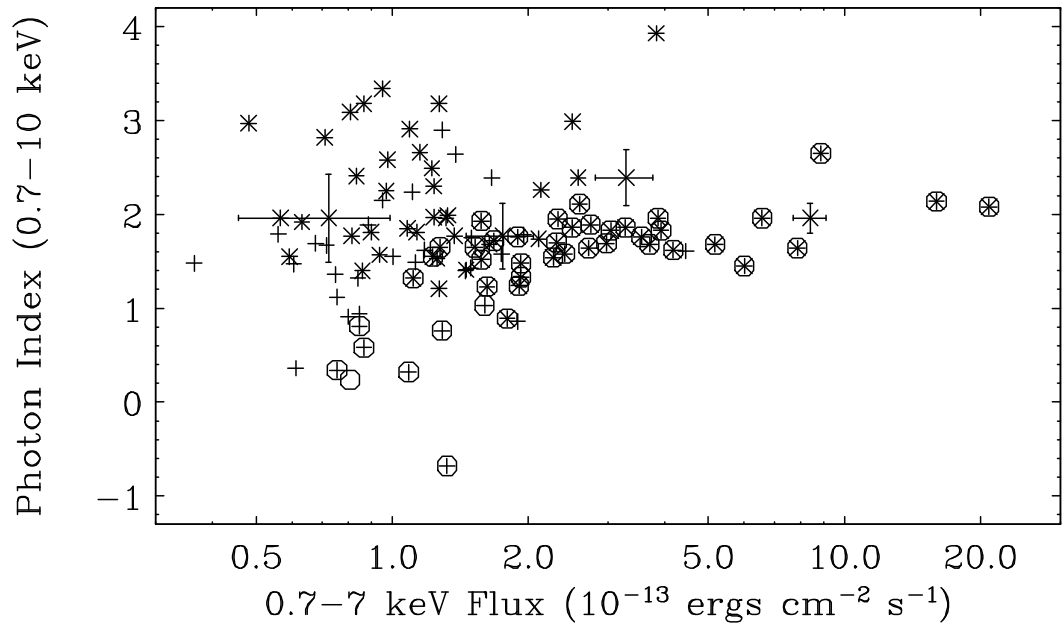}
\caption{
Correlation between the flux and the photon index for all the sources
detected in the LSS field. Crosses, circles, and diagonal crosses
correspond to the detection in the total, hard, and soft band,
respectively. The flux is converted from the GIS count rate in the
0.7--7 keV band assuming a photon index of 1.6. Photon indices are
determined in simultaneous fitting of the GIS and the SIS spectra in
the 0.7--10 keV band with a Galactic absorption fixed at $N_{\rm H}$ =
$1.1\times10^{20}$ cm$^{-2}$. The four large crosses show typical 1
$\sigma$ statistical errors. 
\label{fig2}}
\end{figure}

\placefigure{fig2}

\begin{figure}
\epsscale{0.6}
\plotone{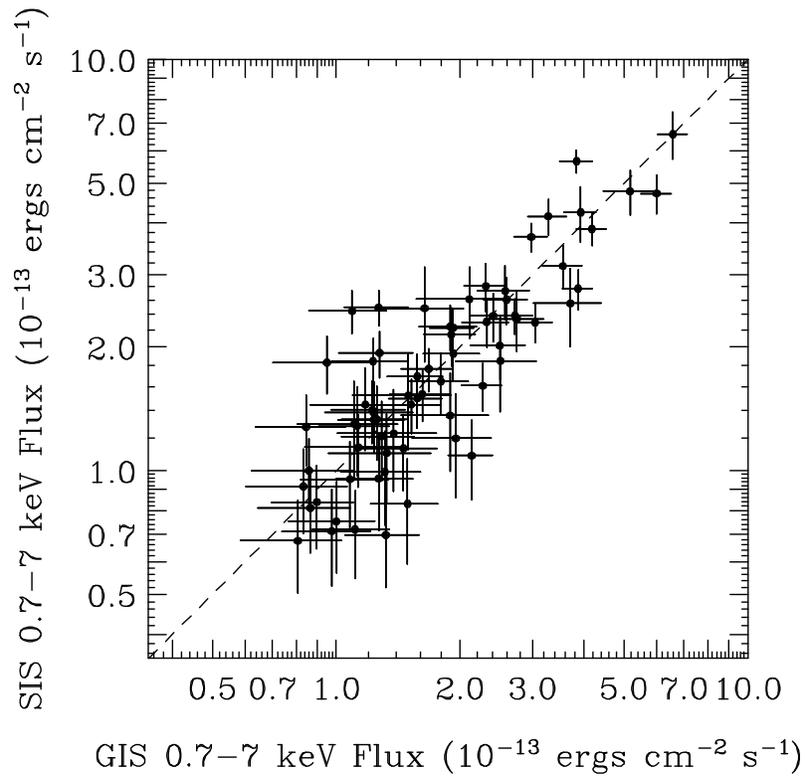}
\caption{
Comparison of fluxes between the GIS and the SIS for sources detected
both with the GIS and the SIS in the total band (0.7--7 keV). The two
fluxes are equal at the dashed line (perfect correlation). A photon
index of 1.6 is assumed for conversion from the count rate to the
flux.  Error bars represent 1$\sigma$ statistical errors.
\label{fig3}}
\end{figure}

\placefigure{fig3}

\begin{figure}
\epsscale{1.0}
\plotone{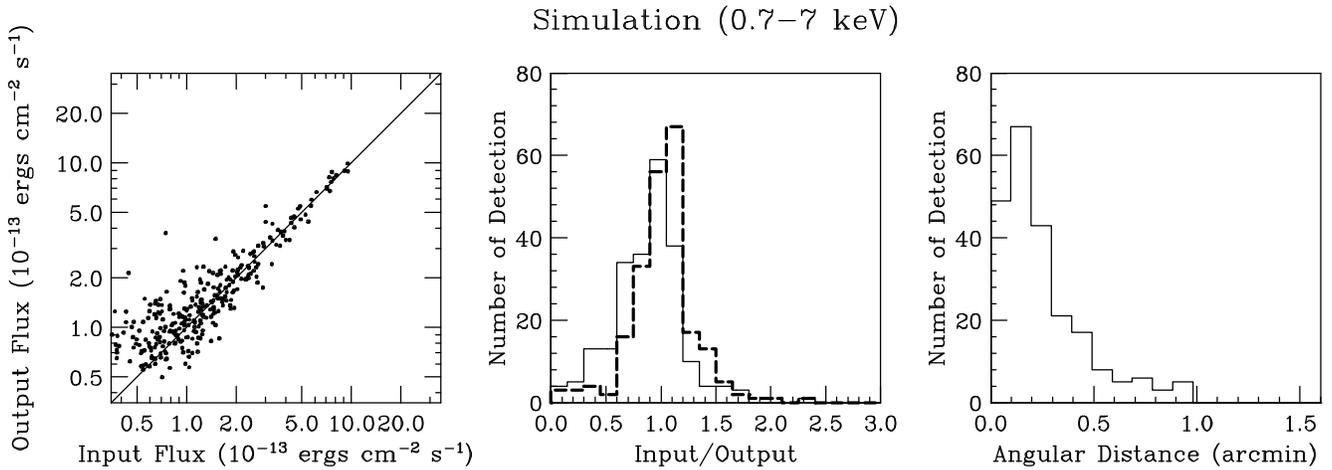}
\caption{
(a) Left: comparison between the input and output flux for simulated data
generated by our instrument simulator. We applied 
the same analysis method as for the real data. 
The input X-ray sky images are created
according to the \logn\ obtained in this paper, extending toward a lower
flux level with a slope of $-3/2$. The input flux is defined as the
flux of the brightest source within a 1 arcmin radius around the
detected position. The output flux equals the input flux at the solid
line.
\newline
(b) Center: histogram of ratio of the input flux to the output flux. The dashed
lines correspond to the case that the input flux is defined as an
integrated (not the brightest) flux within a radius of 1 arcmin around
the detected position.
\newline
(c) Right: histogram of angular distance between the detected position and the
position of the brightest source within a 1 arcmin radius around
the detected position.
\label{fig4}}
\end{figure}

\placefigure{fig4}

\begin{figure}
\epsscale{1.0}
\plotone{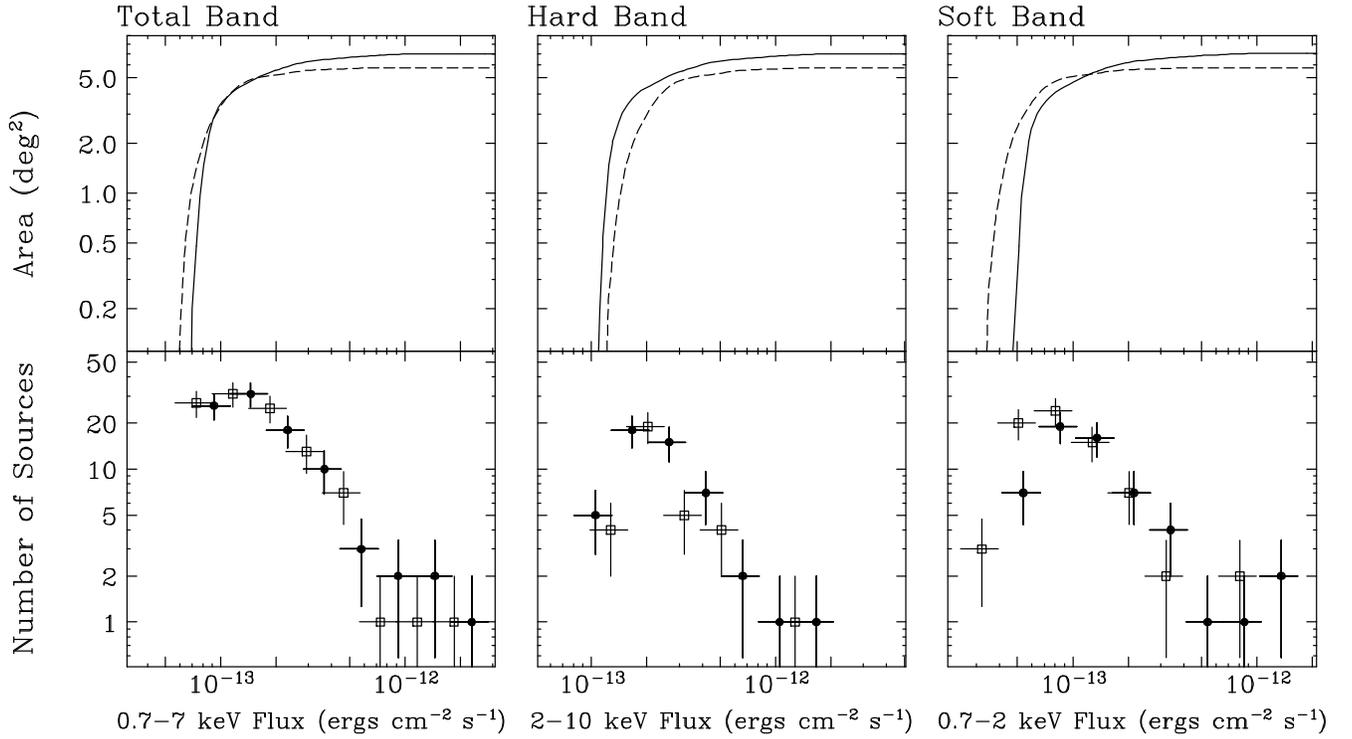}
\caption{
(a) Upper: the observed area, $\Omega$, plotted against the sensitivity limit, 
$S$, in each survey band.  
The solid and dashed lines correspond to the GIS and the SIS, respectively.
The source detection criteria are $\sigma_{\rm g}$ $>3.5$ and
$\sigma_{\rm t}$ $>3.5$ for the GIS, and $\sigma_{\rm s}$ $>3.5$ and
$\sigma_{\rm t}$ $>3.5$ for the SIS. 
A photon index of 1.6 (0.7--7 keV), 1.5 (2--10 keV), and 1.9 (0.7--2
keV) is assumed to convert the count rate to flux.
\newline
(b) Lower: number of sources detected per unit logarithmic flux bin of width 
d log $S$=0.2. The filled circles and open squares correspond to 
the GIS and the SIS, respectively.
\label{fig5}}
\end{figure}

\placefigure{fig5}

\begin{figure}
\epsscale{1.0}
\plotone{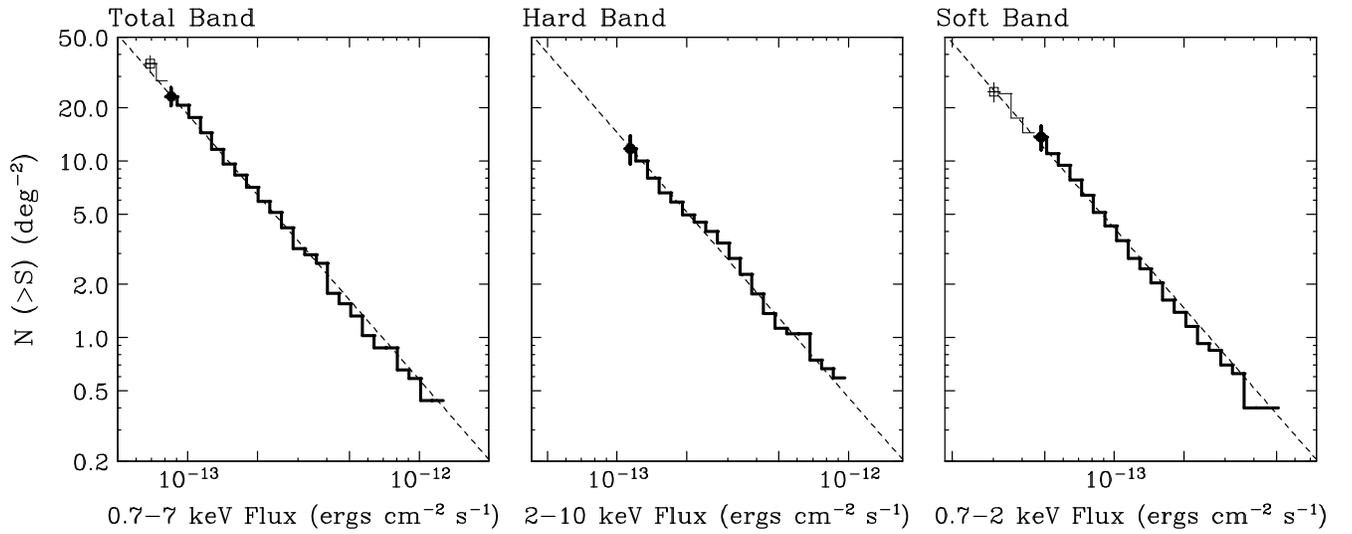}
\caption{
The \logn s derived from the simulated data. The same analysis method
as for the real data is applied. The dashed lines represent the input
\logn s, which is assumed to extend to the lower flux level with a
slope of $-3/2$ until the surface brightness of the CXB is explained. 
In the 0.7--7 keV and 0.7--2 keV band, the steps with thin lines
including the faintest data point (open square)
are determined mainly by the SIS.
\label{fig6}}
\end{figure}

\placefigure{fig6}

\begin{figure}
\epsscale{0.5}
\plotone{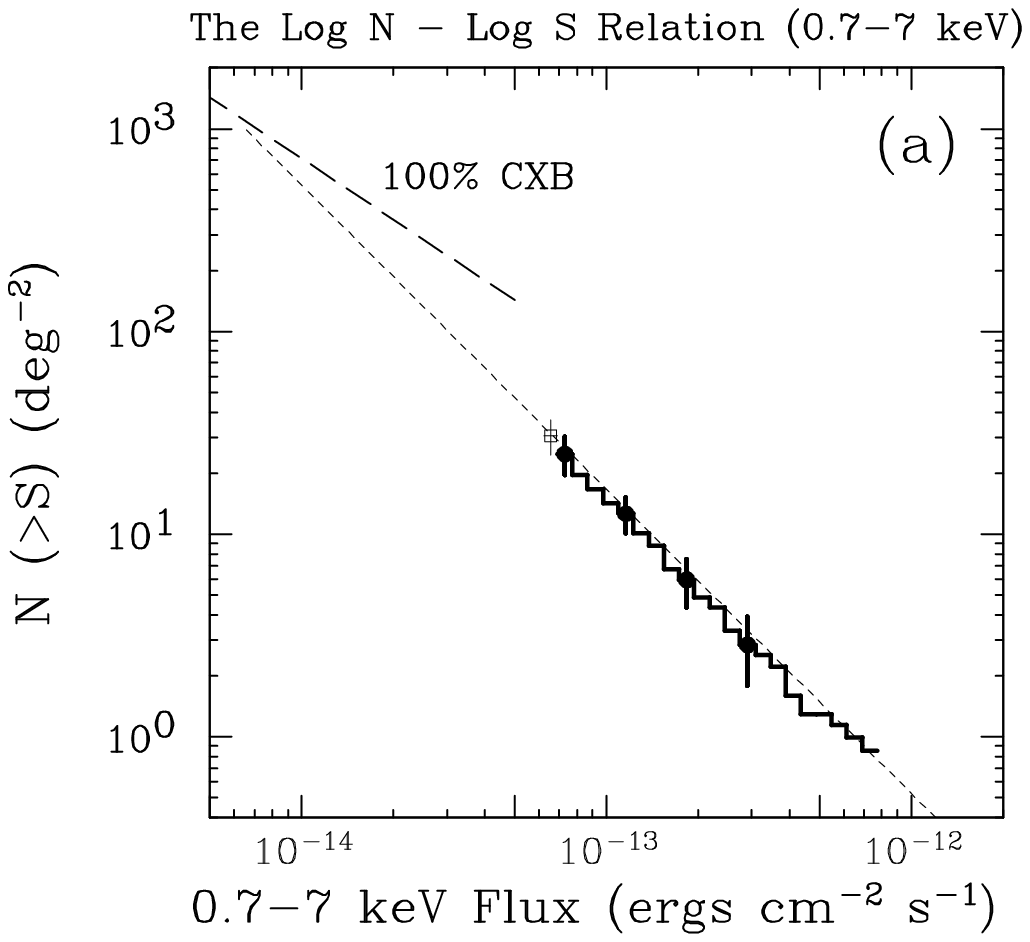}
\plotone{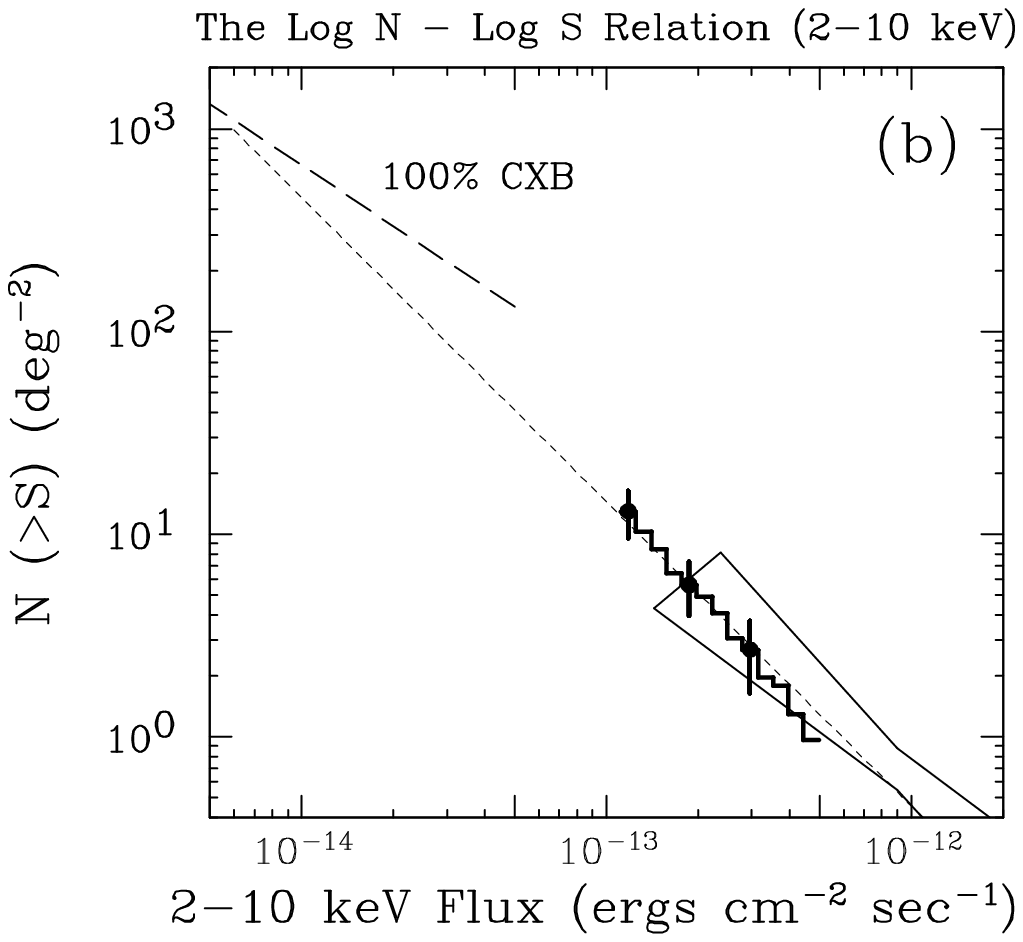}

\bigskip
\plotone{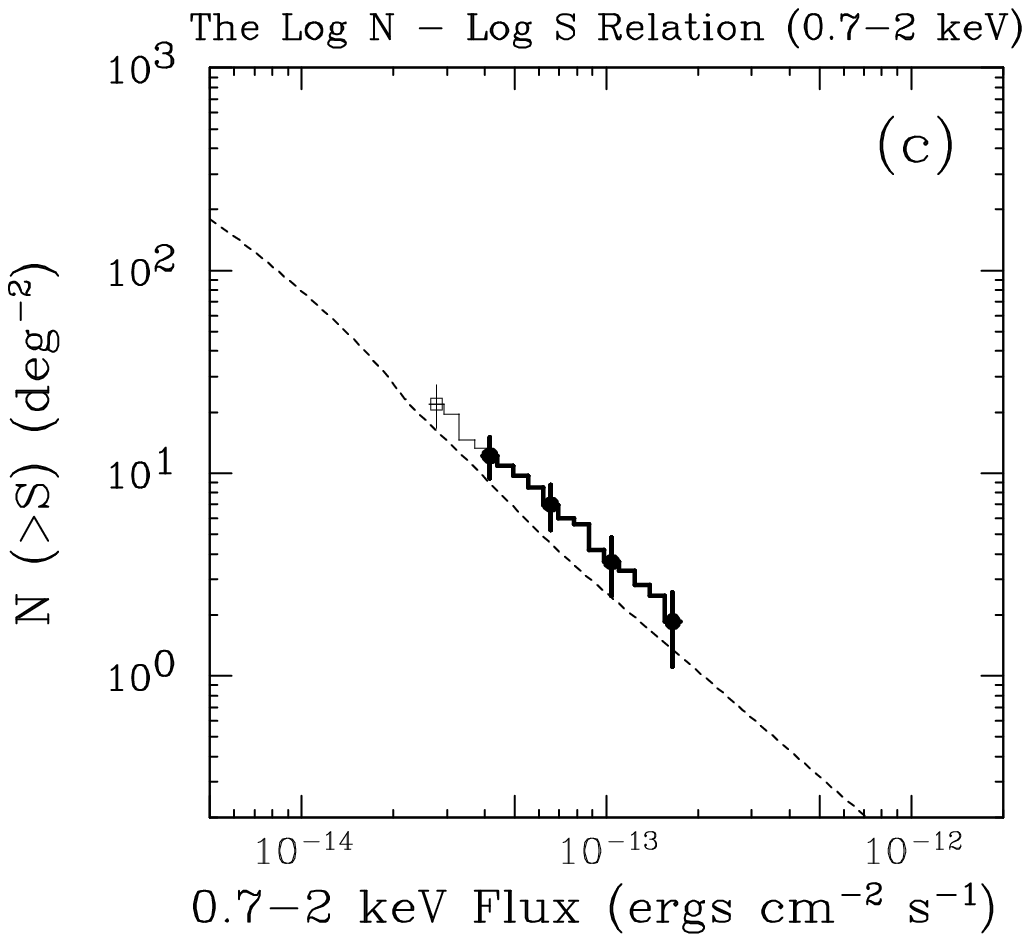}
\caption{\footnotesize
The integral \logn s in the three energy bands.  The steps represent
the present results derived from the LSS.  The associated systematic
errors such as effects of the source confusion were corrected based on
the simulation studies (see text). The error attached at several data
points, marked with open squares (SIS) or filled circles (GIS), shows
a 90\% statistical error coming from the Poisson error in the number
of detected sources. The long-dashed lines in the 0.7--7 keV and the
2--10 keV band correspond to the condition that the total intensity of
the CXB (Ishisaki \etal\ 1999) is explained by the integral of sources
assuming $N(>S)$ is proportional to $S^{-3/2}$. For conversion of the
count rate to the flux, we assume a photon index of 1.6, 1.5, and 1.9,
for the total, hard, and soft bands, respectively. The flux is
corrected for the Galactic absorption to give the unabsorbed (emitted)
flux.
\newline
(a) 0.7--7 keV:
the faintest data point (open square) is determined mainly by
the SIS and the others are by the GIS. The short-dashed line
represents extrapolation with a slope of $-3/2$ from the hard-band
source counts by \heao\ A2 (Piccinotti \etal\ 1982) by converting the
2--10 keV flux to the 0.7--7 keV flux assuming a photon index of
1.6.
\newline
(b) 2--10 keV: 
all the data points are determined mainly by the GIS. The short-dashed
line represents extrapolation with a slope of $-3/2$ from the \heao\
A2 result, whose sensitivity limit is $3\times10^{-11}$
\ergs\ (2--10 keV).  The contour represents the constraints
from the fluctuation analysis by \ginga\ (Hayashida, Inoue, \& Kii
1990; Butcher \etal\ 1997).
\newline
(c) 0.7--2 keV:
the steps with thin lines including the faintest point (open square) 
are determined
mainly by the SIS and the others are by the GIS.  The dashed curve
shows the \rosat\ results, converting the flux in the 0.5--2 keV to
that in the 0.7--2 keV assuming a photon index of 1.9. 
\label{fig7}}
\end{figure}

\placefigure{fig7}

\begin{figure}
\epsscale{0.5}
\plotone{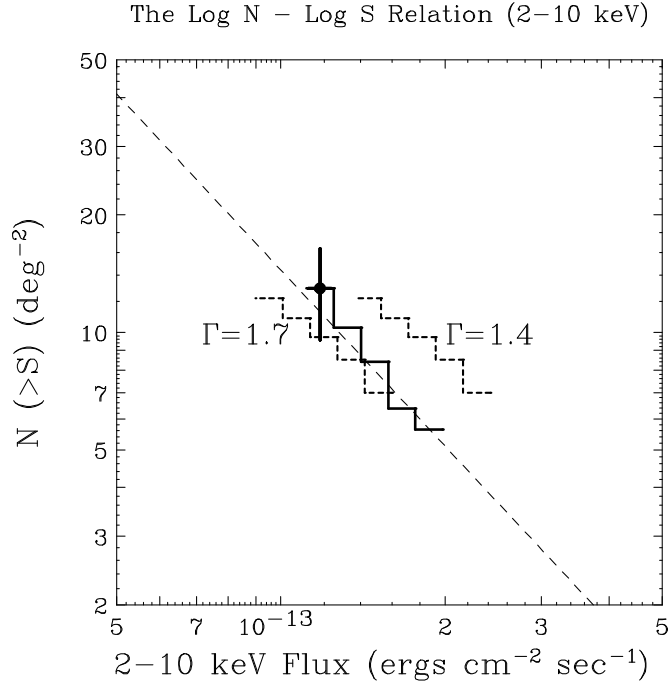}
\caption{
Comparison of the \logn\ between above and below 2 keV 
derived from the LSS.
The steps with thick solid lines represent the \logn\ in the 2--10 keV band.
The steps with dashed lines represent the soft band \logn\ converted to
the hard band flux by assuming two photon indices (1.4 and 1.7).
The long dashed line represents extrapolation with a
slope of $-3/2$ from the \heao\ A2 result.
\label{fig8}}
\end{figure}

\placefigure{fig8}

\begin{figure}
\epsscale{0.5}
\plotone{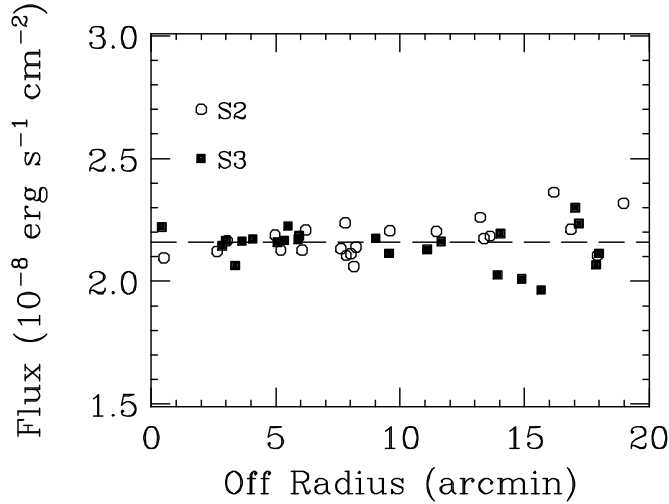}
\caption{
The 2--10 keV fluxes of Crab obtained with the GIS+XRT response that
is used in this paper from the spectra corresponding to different
observing positions, plotted against the radius from the optical axis.
Dead time is corrected. Open and filled circles correspond to GIS2 and
GIS3, respectively.
\label{fig9}}
\end{figure}

\placefigure{fig9}

\clearpage
\begin{deluxetable}{cccc}
\tablenum{1}
\tablewidth{400pt}
\tablecaption{Log of the LSS Observations\label{tbl-1}} 
\tablehead{
\colhead{Start}& \colhead{End} &\colhead{Pointings} &
\colhead{Net Exposure (ksec)\tablenotemark{a}}}
\startdata
1993/12/26 10:22 & 1993/12/27 10:00 &4& 38\nl
1994/01/04 10:32 & 1994/01/07 19:30 &12& 122\nl
1994/06/16 16:36 & 1994/06/19 08:31 &10& 87\nl
1994/06/27 23:12 & 1994/06/30 13:20 &10& 80\nl
1994/12/25 21:35 & 1994/12/26 13:25 &2& 18\nl
1995/01/06 08:16 & 1995/01/06 14:30 &1& 11\nl
1995/01/07 20:55 & 1995/01/08 11:10 &2& 20\nl
1995/01/09 10:54 & 1995/01/13 07:50 &15& 147\nl
1995/06/18 16:23 & 1995/06/20 05:01 &5& 51\nl
1995/07/01 05:03 & 1995/07/02 12:51 &5& 50\nl
1995/07/05 22:13 & 1995/07/08 11:01 &10& 97\nl
\tablenotetext{a}{Based on the data selection for the GIS.}
\enddata
\end{deluxetable}

\begin{deluxetable}{ccccc}
\tablenum{2}
\tablewidth{400pt}
\tablecaption{Number of Detected Sources\label{tbl-2}} 
\tablehead{
\colhead{Detection Band}& \colhead{Total} &\colhead{Hard} &
\colhead{Soft}& \colhead{One Band Only\tablenotemark{a}}}
\startdata
Total&105&43&71&27\nl
Hard&&44&36&1\nl
Soft&&&72&1\nl
\tablenotetext{a}{Number of sources detected only in one band given in the first column}
\enddata
\end{deluxetable}

\begin{deluxetable}{ccccccccccccc}
\scriptsize
\tablenum{3}
\tablewidth{520pt}
\tablecaption{The source list in the LSS field\label{tbl-3}} 
\tablehead{
\colhead{}&
\multicolumn{2}{c}{Position (J2000)\tablenotemark{a}}& \colhead{}&
\multicolumn{3}{c}{Significance (GIS/SIS)\tablenotemark{b}}& \colhead{}&
\multicolumn{3}{c}{Count Rate (GIS/SIS)\tablenotemark{b,c}}& \colhead{}&
\colhead{Index\tablenotemark{d}}\\
\cline{2-3} \cline{5-7} \cline{9-11} \cline{13-13}\\
\colhead{Source Name}&
\colhead{R.A.} &\colhead{DEC.}& \colhead{}&
\colhead{total}&\colhead{hard}&\colhead{soft}& \colhead{}&
\colhead{total}&\colhead{hard}&\colhead{soft}& \colhead{}&
\colhead{$\Gamma$}\\
\colhead{}&
\colhead{}&\colhead{}& \colhead{}&
\colhead{($\sigma$)}&\colhead{($\sigma$)}&\colhead{($\sigma$)}& \colhead{}&
\colhead{(c ksec$^{-1}$)}&\colhead{(c ksec$^{-1}$)}&\colhead{(c ksec$^{-1}$)}& \colhead{}&
\colhead{}	}
\startdata
AX~J130748+2925& 196.9534&29.4328&& 12.0/7.6& 7.0/4.3& 9.9/6.3&& 10.1/17.8& 4.2/5.8& 6.2/11.9&& 1.96$\pm$0.16\nl
AX~J131028+2930& 197.6206&29.5120&& 1.9/5.8& 1.6/3.1& 1.1/5.0&& 0.9/5.3& 0.6/1.9& 0.4/3.4&& 1.55$\pm$0.37\nl
AX~J130803+2950& 197.0127&29.8435&& 5.6/4.9& 4.4/2.4& 3.1/4.4&& 1.9/3.6& 1.2/1.0& 0.7/2.6&& 1.53$\pm$0.29\nl
AX~J130928+2942& 197.3677&29.7066&& 5.1/4.1& 3.3/2.3& 3.5/3.4&& 1.7/3.5& 0.9/1.2& 0.8/2.2&& 1.49$\pm$0.35\nl
AX~J130840+2955& 197.1697&29.9242&& 3.6/3.4& 2.9/3.7& 1.9/0.9&& 1.2/2.2& 0.8/1.9& 0.4/0.4&& 0.34$\pm$0.44\nl
AX~J130926+2952& 197.3605&29.8670&& 4.0/5.1& 3.9/4.0& 1.9/3.3&& 1.3/3.5& 1.0/1.8& 0.4/1.6&& 0.81$\pm$0.30\nl
AX~J131111+2941& 197.7992&29.6880&& 3.7/3.6& 1.1/2.4& 4.2/2.6&& 2.1/3.3& 0.4/1.6& 1.8/1.7&& 2.64$\pm$0.62\nl
AX~J130826+3005& 197.1101&30.0946&& 11.3/6.5& 7.0/3.5& 8.8/5.6&& 6.0/11.5& 2.7/3.7& 3.4/7.8&& 1.83$\pm$0.17\nl
AX~J131220+2938& 198.0847&29.6360&& 7.4/--& 4.9/--& 5.5/--&& 12.1/--& 6.1/--& 6.4/--&& 1.64$\pm$0.29\nl
AX~J130956+2959& 197.4858&29.9883&& 2.9/4.1& 1.2/1.7& 3.5/3.8&& 1.0/2.8& 0.3/0.7& 0.8/2.1&& 1.92$\pm$0.45\nl
AX~J131015+3004& 197.5636&30.0720&& 8.8/7.5& 6.2/3.5& 6.6/6.7&& 3.5/7.6& 2.0/2.0& 1.8/5.6&& 1.70$\pm$0.19\nl
AX~J130851+3015& 197.2166&30.2505&& 4.2/3.9& 3.6/0.9& 2.2/4.1&& 2.5/6.7& 1.8/1.0& 0.9/4.9&& 1.68$\pm$0.41\nl
AX~J131054+3004& 197.7257&30.0689&& 8.7/7.4& 5.2/5.2& 7.3/5.4&& 3.7/6.5& 1.7/3.1& 2.2/3.4&& 1.58$\pm$0.19\nl
AX~J131021+3019& 197.5897&30.3280&& 7.0/6.2& 5.3/4.6& 4.7/4.2&& 2.9/5.8& 1.8/2.9& 1.3/3.0&& 1.24$\pm$0.21\nl
AX~J131204+3007& 198.0169&30.1191&& 3.1/3.5& 1.9/1.8& 2.8/3.2&& 1.4/2.2& 0.6/0.7& 0.9/1.6&& 1.89$\pm$0.49\nl
AX~J131214+3006& 198.0619&30.1071&& 5.3/3.2& 3.9/3.0& 3.1/1.7&& 2.5/2.2& 1.4/1.5& 0.9/0.8&& 1.03$\pm$0.33\nl
AX~J130954+3024& 197.4773&30.4079&& 2.7/3.9& 2.4/3.2& 1.8/2.4&& 0.9/3.0& 0.7/2.0& 0.4/1.2&& 0.36$\pm$0.43\nl
AX~J131158+3010& 197.9952&30.1754&& 3.6/5.1& 2.9/3.1& 2.6/4.0&& 1.3/2.7& 0.8/1.1& 0.7/1.6&& 1.40$\pm$0.33\nl
AX~J131039+3021& 197.6625&30.3585&& 3.8/6.3& 1.0/1.0& 4.1/6.5&& 1.5/5.0& 0.3/0.4& 1.1/4.3&& 3.34$\pm$0.59\nl
AX~J131008+3041& 197.5367&30.6883&& 2.8/4.2& 1.1/3.0& 2.5/2.9&& 1.3/3.0& 0.4/1.6& 0.8/1.5&& 1.32$\pm$0.44\nl
AX~J131205+3031& 198.0209&30.5185&& 6.7/6.5& 3.9/3.1& 5.5/5.8&& 2.4/4.1& 1.0/1.2& 1.5/2.9&& 1.93$\pm$0.26\nl
AX~J130957+3046& 197.4914&30.7812&& 4.2/3.7& 3.8/2.5& 2.2/2.8&& 2.9/3.7& 2.2/1.8& 0.9/1.9&& 0.86$\pm$0.34\nl
AX~J131014+3052& 197.5609&30.8693&& 4.6/3.5& 2.3/2.3& 4.0/2.7&& 3.0/3.2& 1.1/1.4& 1.8/1.9&& 1.78$\pm$0.41\nl
AX~J131112+3049& 197.8005&30.8292&& 11.0/15.8& 2.1/3.2& 12.1/16.0&& 5.9/15.3& 0.6/1.2& 5.2/14.0&& 3.93$\pm$0.34\nl
AX~J131404+3032& 198.5192&30.5481&& 2.1/4.2& 1.4/0.3& 1.7/4.3&& 1.3/4.5& 0.8/0.2& 0.7/3.7&& 3.18$\pm$1.67\nl
AX~J131210+3048& 198.0432&30.8025&& 5.0/4.6& 3.8/3.7& 2.7/2.7&& 2.0/3.3& 1.2/1.9& 0.7/1.4&& 0.76$\pm$0.31\nl
AX~J131255+3045& 198.2321&30.7659&& 3.0/6.1& 2.9/1.9& 2.3/6.6&& 1.5/4.7& 0.8/0.8& 0.8/3.6&& 2.25$\pm$0.40\nl
AX~J131156+3054& 197.9854&30.9120&& 4.8/7.4& 2.5/4.4& 4.0/6.0&& 1.9/5.0& 0.7/1.9& 1.1/3.0&& 1.55$\pm$0.26\nl
AX~J131044+3107& 197.6870&31.1330&& 3.2/3.5& 1.3/1.8& 3.5/3.1&& 2.5/3.8& 0.7/1.1& 1.8/2.7&& 2.39$\pm$0.64\nl
AX~J131128+3105& 197.8696&31.0953&& 4.9/7.7& 4.0/3.6& 2.9/6.9&& 2.0/5.2& 1.3/1.5& 0.8/3.7&& 1.65$\pm$0.24\nl
AX~J131047+3112& 197.6961&31.2098&& 5.3/4.6& 4.8/2.1& 2.7/4.2&& 5.7/6.9& 3.9/1.8& 1.8/5.1&& 1.68$\pm$0.32\nl
AX~J131147+3109& 197.9475&31.1591&& 3.9/5.0& 2.0/2.6& 2.5/4.4&& 1.7/3.1& 0.7/1.0& 0.7/2.1&& 1.81$\pm$0.38\nl
AX~J131321+3100& 198.3401&31.0082&& 6.1/5.8& 6.0/3.4& 2.8/4.8&& 2.8/4.5& 2.3/1.7& 0.8/2.8&& 0.89$\pm$0.23\nl
AX~J131249+3107& 198.2063&31.1195&& 1.4/4.0& 0.0/0.5& 1.2/4.5&& 0.4/2.0& 0.0/0.1& 0.3/1.9&& 3.63$\pm$1.61\nl
AX~J131358+3103& 198.4956&31.0540&& 3.6/4.0& 1.9/1.9& 3.4/3.6&& 2.0/3.0& 0.8/0.8& 1.4/2.0&& 1.99$\pm$0.49\nl
AX~J131249+3112& 198.2056&31.2165&& 12.1/11.3& 8.2/6.3& 8.8/9.4&& 6.4/10.5& 3.3/3.6& 3.3/6.9&& 1.62$\pm$0.13\nl
AX~J131211+3127& 198.0468&31.4515&& 6.9/6.7& 3.2/2.7& 5.9/6.2&& 4.0/7.4& 1.3/1.7& 2.4/5.7&& 2.39$\pm$0.30\nl
AX~J131321+3119& 198.3378&31.3250&& 8.9/7.5& 6.8/3.6& 6.1/6.7&& 3.5/4.4& 2.1/1.2& 1.6/3.1&& 1.54$\pm$0.18\nl
AX~J131345+3118& 198.4385&31.3063&& 10.8/12.6& 6.8/7.1& 8.9/10.3&& 4.6/10.0& 2.2/3.5& 2.7/6.5&& 1.69$\pm$0.14\nl
AX~J131609+3105& 199.0409&31.0868&& 5.7/--& 3.7/--& 4.1/--&& 6.9/--& 3.5/--& 3.2/--&& 1.61$\pm$0.38\nl
AX~J131512+3114& 198.8026&31.2444&& 2.6/3.7& 1.4/2.7& 2.6/2.6&& 0.9/1.9& 0.4/0.9& 0.7/1.0&& 1.47$\pm$0.45\nl
AX~J131356+3127& 198.4838&31.4625&& 5.9/3.2& 3.2/2.5& 5.1/2.1&& 2.1/1.4& 0.8/0.7& 1.3/0.7&& 1.77$\pm$0.35\nl
AX~J131529+3117& 198.8709&31.2924&& 8.3/7.7& 4.1/4.0& 7.8/6.5&& 4.0/7.1& 1.4/2.2& 2.8/4.8&& 2.11$\pm$0.23\nl
AX~J131444+3123& 198.6853&31.3891&& 5.7/10.2& 1.5/2.5& 6.3/10.3&& 1.9/6.8& 0.3/0.8& 1.7/6.0&& 3.18$\pm$0.43\nl
AX~J131325+3135& 198.3579&31.5953&& 4.7/8.3& 0.8/2.7& 5.0/8.2&& 1.7/6.6& 0.2/1.1& 1.4/5.6&& 2.91$\pm$0.38\nl
AX~J131501+3141\tablenotemark{e}& 198.7575&31.6903&& 4.9/3.9& 5.8/4.5& 1.1/0.5&& 2.0/1.9& 2.0/1.7& 0.2/0.1&& -0.68$\pm$0.45\nl
AX~J131556+3135& 198.9855&31.5933&& 2.7/4.9& 1.0/2.4& 2.6/4.4&& 0.9/2.5& 0.2/0.7& 0.6/1.7&& 1.96$\pm$0.47\nl
AX~J131327+3155& 198.3634&31.9237&& 9.1/8.4& 6.3/4.1& 6.7/7.5&& 5.5/8.5& 2.9/2.3& 2.7/6.4&& 1.76$\pm$0.17\nl
AX~J131412+3151& 198.5506&31.8631&& 2.7/6.0& 1.8/3.2& 1.5/5.0&& 1.4/5.4& 0.7/1.9& 0.6/3.5&& 1.57$\pm$0.35\nl
AX~J131651+3133& 199.2156&31.5611&& 3.7/3.8& 2.5/1.9& 2.8/3.3&& 2.3/4.1& 1.2/1.2& 1.2/2.9&& 1.75$\pm$0.41\nl
AX~J131414+3153& 198.5616&31.8883&& 3.3/3.7& 0.9/1.3& 3.5/3.7&& 1.8/3.0& 0.3/0.7& 1.4/2.3&& 2.66$\pm$0.75\nl
AX~J131407+3158& 198.5302&31.9750&& 6.3/7.8& 3.8/4.3& 5.5/6.6&& 2.9/6.1& 1.3/2.1& 1.9/4.0&& 1.76$\pm$0.23\nl
AX~J131328+3204& 198.3688&32.0739&& 2.6/5.4& 1.6/2.9& 2.3/4.6&& 1.2/4.7& 0.6/1.5& 0.8/3.3&& 1.77$\pm$0.38\nl
AX~J131408+3202& 198.5350&32.0440&& 4.1/5.3& 1.8/3.2& 4.2/4.2&& 1.9/3.8& 0.6/1.4& 1.4/2.4&& 1.97$\pm$0.37\nl
AX~J131355+3205& 198.4801&32.0962&& 3.9/5.1& 2.6/3.2& 2.8/4.0&& 3.2/7.1& 1.5/2.6& 1.8/4.5&& 1.74$\pm$0.32\nl
AX~J131354+3207& 198.4769&32.1181&& 4.5/4.0& 1.4/0.7& 4.3/4.4&& 3.9/5.0& 0.8/0.4& 2.9/4.7&& 2.99$\pm$0.71\nl
AX~J131512+3157& 198.8022&31.9593&& 3.8/4.4& 2.2/3.0& 2.9/3.3&& 1.8/3.9& 0.8/1.6& 1.0/2.3&& 1.62$\pm$0.37\nl
AX~J131521+3159& 198.8392&31.9871&& 9.3/10.0& 6.2/4.7& 7.3/8.8&& 5.0/11.3& 2.5/3.0& 2.8/8.3&& 1.86$\pm$0.16\nl
AX~J131639+3149& 199.1634&31.8274&& 6.9/8.3& 4.3/4.6& 5.1/6.9&& 2.6/4.8& 1.2/1.6& 1.3/3.2&& 1.72$\pm$0.21\nl
AX~J131511+3201& 198.7981&32.0328&& 5.3/3.5& 3.4/1.9& 3.4/3.0&& 2.3/2.2& 1.1/0.7& 1.0/1.6&& 1.43$\pm$0.40\nl
AX~J131651+3155& 199.2147&31.9196&& 4.7/4.2& 3.7/3.0& 4.1/3.0&& 1.7/1.9& 1.1/0.9& 1.1/1.0&& 1.32$\pm$0.30\nl
AX~J131514+3208& 198.8085&32.1451&& 4.1/3.8& 1.3/1.8& 4.5/3.4&& 1.5/1.9& 0.3/0.5& 1.2/1.4&& 2.58$\pm$0.57\nl
AX~J131709+3154& 199.2908&31.9075&& 4.6/3.9& 3.9/0.5& 2.2/4.1&& 2.0/2.7& 1.2/0.2& 0.6/2.4&& 1.96$\pm$0.43\nl
AX~J131539+3206& 198.9138&32.1139&& 1.7/4.3& 2.2/2.2& 0.9/3.8&& 0.6/2.5& 0.6/0.8& 0.2/1.7&& 1.48$\pm$0.44\nl
\nl\nl\nl\nl
AX~J131742+3152& 199.4286&31.8801&& 6.0/6.0& 3.8/2.8& 4.8/5.4&& 4.2/6.3& 2.0/1.7& 2.3/4.7&& 1.89$\pm$0.27\nl
AX~J131522+3218& 198.8457&32.3056&& 4.5/4.4& 3.0/1.8& 3.5/4.1&& 1.4/2.3& 0.7/0.6& 0.8/1.7&& 1.81$\pm$0.37\nl
AX~J131724+3203& 199.3513&32.0567&& 6.0/7.0& 4.1/4.9& 4.1/5.0&& 2.5/4.2& 1.3/1.9& 1.2/2.3&& 1.23$\pm$0.22\nl
AX~J131638+3211& 199.1605&32.1919&& 3.3/4.1& 2.1/2.4& 2.9/3.4&& 1.1/1.9& 0.5/0.7& 0.7/1.2&& 1.67$\pm$0.42\nl
AX~J131650+3222& 199.2123&32.3812&& 3.3/4.4& 2.5/2.2& 2.0/3.9&& 1.0/2.4& 0.6/0.7& 0.4/1.7&& 1.69$\pm$0.41\nl
AX~J131526+3234& 198.8585&32.5801&& 3.6/4.3& 1.8/1.7& 3.4/4.2&& 1.3/2.5& 0.5/0.5& 0.9/2.0&& 2.41$\pm$0.47\nl
AX~J131742+3220& 199.4252&32.3458&& 3.4/2.6& 4.0/2.8& 0.0/0.9&& 1.2/1.3& 1.2/1.0& 0.0/0.3&& 0.24$\pm$0.54\nl
AX~J131730+3222& 199.3784&32.3760&& 8.1/4.5& 4.5/0.8& 7.1/4.8&& 3.3/3.0& 1.3/0.3& 1.9/2.6&& 2.26$\pm$0.30\nl
AX~J131551+3237& 198.9630&32.6262&& 3.4/5.5& 3.5/5.5& 1.3/2.1&& 1.7/3.1& 1.5/2.3& 0.4/0.8&& 0.32$\pm$0.34\nl
AX~J131735+3225& 199.3992&32.4323&& 3.5/3.3& 2.8/2.3& 2.7/2.4&& 1.1/1.7& 0.7/0.8& 0.6/1.0&& 1.36$\pm$0.40\nl
AX~J131707+3237& 199.2825&32.6229&& 7.8/7.6& 3.9/3.9& 6.1/6.5&& 3.6/6.2& 1.3/1.9& 2.0/4.4&& 1.95$\pm$0.24\nl
AX~J131821+3232& 199.5884&32.5385&& 5.5/6.0& 3.7/1.8& 3.7/6.1&& 1.9/3.6& 0.9/0.5& 0.9/3.2&& 2.30$\pm$0.34\nl
AX~J131621+3248& 199.0877&32.8156&& 3.1/5.5& 0.0/2.5& 3.6/4.9&& 1.1/3.4& 0.0/0.9& 1.0/2.5&& 2.82$\pm$0.53\nl
AX~J131816+3240& 199.5670&32.6823&& 8.1/9.3& 6.2/5.5& 5.2/7.5&& 3.0/6.0& 1.8/2.3& 1.3/3.7&& 1.33$\pm$0.16\nl
AX~J131918+3238& 199.8275&32.6453&& 4.1/4.2& 1.7/3.0& 4.0/3.0&& 1.7/2.6& 0.5/1.3& 1.2/1.3&& 1.85$\pm$0.42\nl
AX~J131725+3300& 199.3566&33.0009&& 6.5/7.6& 5.2/4.4& 4.1/6.2&& 2.4/4.6& 1.5/1.6& 1.0/3.0&& 1.52$\pm$0.21\nl
AX~J131654+3304& 199.2280&33.0792&& 4.7/4.0& 2.2/2.6& 3.6/3.3&& 1.9/2.6& 1.0/1.2& 1.0/1.4&& 1.21$\pm$0.32\nl
AX~J131758+3257& 199.4953&32.9635&& 9.7/10.2& 6.6/5.5& 7.0/8.6&& 4.2/6.5& 2.2/2.1& 2.1/4.4&& 1.64$\pm$0.15\nl
AX~J131928+3251& 199.8682&32.8509&& 5.6/6.6& 3.5/3.6& 4.0/5.5&& 2.3/3.9& 1.1/1.3& 1.1/2.6&& 1.65$\pm$0.26\nl
AX~J131813+3301& 199.5571&33.0204&& 3.1/4.0& 2.8/3.0& 1.5/2.7&& 1.3/2.3& 0.9/1.1& 0.4/1.1&& 0.94$\pm$0.38\nl
AX~J131832+3259& 199.6341&32.9873&& 3.9/4.5& 3.2/4.1& 2.1/2.2&& 1.3/2.2& 0.9/1.4& 0.5/0.8&& 0.58$\pm$0.35\nl
AX~J131812+3303& 199.5520&33.0533&& 2.2/4.0& 0.7/2.2& 1.9/3.3&& 0.9/2.1& 0.2/0.7& 0.5/1.4&& 1.79$\pm$0.54\nl
AX~J131844+3306& 199.6835&33.1012&& 3.6/3.9& 1.8/2.1& 4.0/3.3&& 1.2/1.8& 0.4/0.6& 1.0/1.2&& 3.09$\pm$0.50\nl
AX~J131905+3303& 199.7725&33.0624&& 3.2/4.0& 2.7/2.0& 1.2/3.4&& 1.2/1.9& 0.8/0.6& 0.3/1.4&& 1.12$\pm$0.28\nl
AX~J132011+3257& 200.0470&32.9570&& 3.6/3.7& 0.8/2.1& 3.6/3.1&& 1.7/3.5& 0.2/1.2& 1.3/2.3&& 2.24$\pm$0.62\nl
AX~J131904+3308& 199.7685&33.1439&& 4.2/4.0& 3.4/2.2& 2.8/3.3&& 1.5/2.0& 1.1/0.7& 0.7/1.3&& 1.55$\pm$0.22\nl
AX~J131842+3311& 199.6771&33.1966&& 3.6/2.9& 2.4/0.0& 2.7/3.3&& 1.2/1.3& 0.6/0.0& 0.6/1.1&& 0.91$\pm$0.31\nl
AX~J131831+3320& 199.6322&33.3417&& 10.2/9.6& 6.8/5.2& 6.5/8.1&& 4.7/6.2& 2.5/2.0& 2.0/4.2&& 1.83$\pm$0.11\nl
AX~J131850+3326& 199.7111&33.4361&& 20.4/25.8& 8.8/9.4& 18.6/24.2&& 13.6/31.7& 3.7/5.4& 9.7/26.2&& 2.65$\pm$0.09\nl
AX~J132104+3309& 200.2701&33.1590&& 3.7/3.2& 1.1/1.5& 3.9/2.9&& 2.0/2.7& 0.4/0.7& 1.6/2.1&& 2.90$\pm$0.81\nl
AX~J131808+3335& 199.5359&33.5886&& 6.5/5.7& 4.3/2.8& 5.2/5.1&& 3.8/5.5& 1.9/1.6& 2.2/3.9&& 1.86$\pm$0.26\nl
AX~J131928+3330& 199.8693&33.5120&& 4.9/5.8& 2.5/2.0& 4.9/5.7&& 1.9/3.8& 0.7/0.7& 1.4/3.1&& 2.49$\pm$0.39\nl
AX~J132032+3326& 200.1339&33.4426&& 12.1/8.9& 7.3/4.5& 9.7/7.8&& 6.0/7.5& 2.6/2.1& 3.4/5.4&& 1.96$\pm$0.16\nl
AX~J131831+3341& 199.6303&33.6960&& 11.9/9.4& 8.5/5.7& 8.4/7.5&& 9.2/12.8& 5.0/4.9& 4.6/7.9&& 1.45$\pm$0.14\nl
AX~J132054+3328& 200.2264&33.4686&& 1.4/5.1& 0.0/1.6& 1.4/5.2&& 0.7/4.3& 0.0/0.6& 0.6/3.9&& 2.97$\pm$0.70\nl
AX~J131805+3349& 199.5241&33.8247&& 7.1/8.0& 4.2/5.5& 5.0/6.0&& 8.0/13.0& 3.5/5.3& 4.4/7.6&& 1.68$\pm$0.21\nl
AX~J131935+3338& 199.8992&33.6408&& 3.8/3.2& 2.9/0.3& 2.1/3.6&& 1.5/1.7& 0.9/0.1& 0.5/1.6&& 2.15$\pm$0.57\nl
AX~J131822+3347& 199.5955&33.7949&& 25.4/21.6& 15.0/10.0& 20.4/19.3&& 32.1/48.9& 13.0/11.8& 19.2/36.9&& 2.08$\pm$0.07\nl
AX~J131927+3343& 199.8631&33.7201&& 6.5/7.3& 4.9/4.0& 4.5/6.1&& 3.0/5.2& 1.8/1.8& 1.4/3.4&& 1.48$\pm$0.21\nl
AX~J131917+3345& 199.8236&33.7600&& 4.9/4.8& 3.4/2.8& 3.4/3.8&& 2.2/3.1& 1.2/1.2& 1.1/1.8&& 1.41$\pm$0.31\nl
AX~J132010+3351& 200.0443&33.8616&& 4.9/2.3& 3.9/1.1& 3.2/2.0&& 2.7/1.7& 1.7/0.5& 1.2/1.2&& 1.58$\pm$0.37\nl
AX~J131939+3355& 199.9143&33.9307&& 4.0/3.1& 3.9/0.7& 2.1/3.2&& 2.2/2.7& 1.8/0.3& 0.8/2.2&& 1.40$\pm$0.40\nl
AX~J132055+3354& 200.2295&33.9040&& 11.8/--& 6.8/--& 9.8/--&& 24.6/--& 9.7/--& 15.3/--&& 2.14$\pm$0.20\nl
\tablenotetext{a}{Typical position uncertainty is about 0.7 arcmin in radius (see text).}
\tablenotetext{b}{``--'' means out of FOV (only for the SIS)}
\tablenotetext{c}{Vignetting-corrected count rate at the reference position in 
the XRT coordinate ($8'.5$ offset for the GIS and $7'.0$ offset for the SIS)
integrated within a raduis of 6$'$ (GIS) and 4$'$ (SIS).}
\tablenotetext{d}{The best-fit photon index obtained by the simultaneous 
fitting of the GIS and the SIS spectra in the 0.7--10 keV range, 
assuming a power law 
with the Galactic absorption ($1.1 \times 10^{20}$ cm$^{-2}$). Error is a 
$1 \sigma$ statistical error for a sigle parameter. }
\tablenotetext{e}{Corresponds to the LSS hardest source. The position is taken from results of deep X-ray
observations by Sakano \etal\ (1998). }
\enddata
\end{deluxetable}

\begin{deluxetable}{ccccccc}
\tablenum{4}
\tablewidth{400pt}
\tablecaption{Conversion Factors from Count Rate to Flux\label{tbl-4}} 
\tablehead{
\colhead{Assumed Spectrum\tablenotemark{a}}& 
\multicolumn{6}{c}{Conversion Factor from Count Rate to Flux\tablenotemark{b}}\\
\colhead{}& 
\multicolumn{2}{c}{Total Band\tablenotemark{c}}&
\multicolumn{2}{c}{Hard Band\tablenotemark{d}}&
\multicolumn{2}{c}{Soft Band\tablenotemark{e}}\\
\colhead{Photon Index}&
\colhead{GIS}&\colhead{SIS}&
\colhead{GIS}&\colhead{SIS}&
\colhead{GIS}&\colhead{SIS}}
\startdata
1.0&0.746&0.478&1.382&1.226&0.408&0.192 \nl
1.4&0.678&0.401 &1.201&1.033 &0.431&0.194 \nl
1.5&0.664&0.385 &1.161&0.992 &0.438&0.194 \nl
1.6&0.650&0.369 &1.122&0.954 &0.444&0.194 \nl
1.7&0.638&0.355 &1.086&0.918 &0.451&0.195 \nl
1.8&0.628&0.341 &1.051&0.884 &0.458&0.195 \nl
1.9&0.618&0.329 &1.018&0.853 &0.465&0.196 \nl
2.0&0.610&0.317 &0.986&0.823 &0.472&0.196 \nl
3.0&0.592&0.246 &0.751&0.615 &0.561&0.202 \nl
\tablenotetext{a}{A Galactic absorption of $N_{\rm H}=1.1\times10^{20}$ cm$^{-2}$ is assumed.}
\tablenotetext{b}{In units of [10$^{-13}$ \ergs ] / [count ksec$^{-1}$]. 
The count rate is the vignetting-corrected count rate given in the source list
(Table 3). Flux is corrected for the Galactic absorption (i.e., unabsorbed flux).}
\tablenotetext{c}{From 0.7--7 keV count rate to 0.7--7 keV flux}
\tablenotetext{d}{From 2--10 keV count rate (2--7 keV for SIS) to 2--10 keV flux}
\tablenotetext{e}{From 0.7--2 keV count rate to 0.7--2 keV flux}
\enddata
\end{deluxetable}

\begin{deluxetable}{clcccc}
\tablenum{5}
\tablewidth{400pt}
\tablecaption{Average Spectra for Flux Limited Samples\label{tbl-5}} 
\tablehead{
\multicolumn{3}{c}{Sample}&\colhead{}&
\multicolumn{2}{c}{Best Fit Photon Index\tablenotemark{a}}\\
\cline{1-3} \cline{5-6}\\
\colhead{Survey Band}& \colhead{Flux Range} &\colhead{Number of}&\colhead{}&
\multicolumn{2}{c}{Fitting Range}\\
\colhead{}&\colhead{($10^{-13}$ erg s$^{-1}$ cm$^{-2}$)}&\colhead{Sources}&\colhead{}&
\colhead{0.7--10 keV}&\colhead{2--10 keV}}
\startdata
Total& 0.5--2.0 (0.7--7 keV)&74&\colhead{}&
1.63$\pm$0.07&1.63$\pm$0.18\nl
Hard& 0.8--4.0 (2--10 keV)&36&\colhead{}&
(1.51$\pm$0.05)\tablenotemark{b}&1.49$\pm$0.10\nl
Soft& 0.2--2.3 (0.7--2 keV)&64&\colhead{}&
(1.76$\pm$0.07)\tablenotemark{b}&1.85$\pm$0.22\nl
\tablenotetext{a}{A Galactic absorption of $N_{\rm H}=1.1\times10^{20}$ cm$^{-2}$ is assumed. The error indicates 1 $\sigma$ statistical error for a single parameter.}
\tablenotetext{b}{This value could be affected by statistical bias (see text).}
\enddata
\end{deluxetable}

\begin{deluxetable}{ccccc}
\tablenum{6}
\tablewidth{400pt}
\tablecaption{Estimation of Number Fraction of Populations in Our Sample\label{tbl-6}} 
\tablehead{
\colhead{Population}&
\colhead{Assumed Spectrum}&
\multicolumn{3}{c}{Expected Percentage in the Sample\tablenotemark{a}}\\
\colhead{}&\colhead{}&
\colhead{0.7--7 keV}&\colhead{2--10 keV}&\colhead{0.7--2 keV\tablenotemark{b}}}
\startdata
AGN& $\Gamma=2.0$\tablenotemark{c}&45 (65)&30 (40)&60 (80)\nl
   & $\Gamma=1.6$\tablenotemark{c}&55 (75)&50 (70)&60 (80)\nl
CG & $kT=6$\tablenotemark{d}&10 &10&10\nl
Star&$\Gamma=3.0$\tablenotemark{c}&5--10&$<$2&10--15\nl
Other\tablenotemark{e}&	&25--40 (5--20)&50--60 (20--50)&15--20 (0)\nl
\tablenotetext{a}{The parenthesis indicates an extreme case under the possible
uncertainty in the identifications of the \rosat\ sources (see text).}
\tablenotetext{b}{The estimation in the soft band is made based on the results
by \rosat\ in the 0.5--2 keV band (Hasinger 1996 and references therein).}
\tablenotetext{c}{Power law photon index.}
\tablenotetext{d}{Temperature in keV for a Raymond-Smith plasma model with an
elemental abundance of 0.3 solar value and a redshift of 0.2.}
\tablenotetext{e}{
Includes NELGs and unidentified objects in the \rosat\ surveys.}
\enddata
\end{deluxetable}

\end{document}